\documentclass[english]{IEEEtran}
\usepackage[T1]{fontenc}
\usepackage[latin9]{inputenc}
\usepackage{xcolor}
\usepackage{pdfcolmk}
\usepackage{amsthm}
\usepackage{amsmath}
\usepackage{amssymb}
\usepackage{graphicx}
\PassOptionsToPackage{normalem}{ulem}
\usepackage{ulem}

\makeatletter

\providecolor{lyxadded}{rgb}{0,0,1}
\providecolor{lyxdeleted}{rgb}{1,0,0}

  \theoremstyle{plain}
  \newtheorem{lem}{\protect\lemmaname}
  \theoremstyle{plain}
  \newtheorem{thm}{\protect\theoremname}
  \theoremstyle{plain}
  \newtheorem{cor}{\protect\corollaryname}
  \theoremstyle{definition}
  \newtheorem{defn}{\protect\definitionname}

\usepackage{cite}
\usepackage{times}

\newtheorem{assumption}{Assumption}

\newtheorem{condition}{Condition}
\usepackage{subfig}

\makeatother

\usepackage{babel}
\providecommand{\corollaryname}{Corollary}
\providecommand{\definitionname}{Definition}
\providecommand{\lemmaname}{Lemma}
\providecommand{\theoremname}{Theorem}

\begin{document}

\title{How Much Cache is Needed to Achieve Linear Capacity Scaling in Backhaul-Limited
Dense Wireless Networks? }

\author{{\normalsize{}An Liu, }\textit{\normalsize{}Member IEEE}{\normalsize{},
and Vincent Lau,}\textit{\normalsize{} Fellow IEEE}{\normalsize{},\\Department
of Electronic and Computer Engineering, Hong Kong University of Science
and Technology}%
\thanks{This work was supported by RGC16204814.%
}\vspace{-0.3in}
}
\maketitle
\begin{abstract}
Dense wireless networks are a promising solution to meet the huge
capacity demand in 5G wireless systems. However, there are two implementation
issues, namely the interference and backhaul issues. To resolve these
issues, we propose a novel network architecture called the \textit{backhaul-limited
cached dense wireless network (C-DWN)}, where a \textit{physical layer
(PHY) caching} scheme is employed at the base stations (BSs) but only
a fraction of the BSs have wired payload backhauls. The PHY caching
can replace the role of wired backhauls to achieve both the \textit{cache-induced
MIMO cooperation gain} and \textit{cache-assisted Multihopping gain}.
Two fundamental questions are addressed. Can we exploit the PHY caching
to achieve linear capacity scaling with limited payload backhauls?
If so, how much cache is needed? We show that the capacity of the
backhaul-limited C-DWN indeed scales linearly with the number of BSs
if the BS cache size is larger than a threshold that depends on the
content popularity. We also quantify the throughput gain due to cache-induced
MIMO cooperation over conventional caching schemes (which exploit
purely the cached-assisted multihopping). Interestingly, the minimum
BS cache size needed to achieve a significant cache-induced MIMO cooperation
gain is the same as that needed to achieve the linear capacity scaling. \end{abstract}


\section{Introduction}

In dense wireless networks, the dense BS deployment brings the network
closer to mobile users and thus can significantly improve spectral
efficiency per unit area. For instance, the capacity of dense wireless
networks scales linearly with the number of BSs $N$ if all BSs are
equipped with wired payload backhauls%
\footnote{This linear capacity scaling law may be violated if we allow Device-to-Device
(D2D) communications between users. We leave the consideration of
D2D as future work.%
}. However, there are also two key technical challenges: the \textit{interference}
and the \textit{backhaul cost issue}s. 

Compared to traditional cellular networks, the interference in dense
wireless networks is more severe due to the increased BS density.
Recently, some advanced interference mitigation schemes such as cooperative
MIMO (Co-MIMO) \cite{zhang2004cochannel,somekh2009cooperative,Irmer_Comm11_CoMPsurvey}
have been proposed. By sharing both channel state information (CSI)
and payload data among the concerned BSs, the Co-MIMO can transform
the wireless network from an unfavorable interference topology to
a favorable broadcast topology, where the interference can be mitigated
much more efficiently. However, the Co-MIMO technique requires high
capacity backhaul for payload exchange between BSs, which is a cost
bottleneck in dense wireless networks. 

Moreover, a dense wireless network with all small BSs having wired
payload backhauls suffers from high CAPEX and OPEX \cite{Paolini_mobiledatabackhaul}.
To address this issue, a more flexible backhaul solution has been
proposed where only a fraction of $N_{0}\ll N$ BSs have wired payload
backhauls, and the other BSs are connected to the core network via
low-cost wireless backhauls \cite{Caire_INFOCOM12_femtocache,Shi_ICSPCC2014_wirelessBH,Ge_Network2014_wirelessBH,Ericsson_review_wirelessBH}.
However, in this case, the performance is limited by the number of
wired payload backhauls $N_{0}$ and the total network capacity can
only scale as $\Theta\left(N_{0}\right)$. 

This paper addresses the following fundamental questions in dense
wireless networks. Can we realize the benefit of MIMO cooperation
without wired backhaul connections between the BSs? Can we achieve
linear capacity scaling with only $N_{0}\ll N$ wired payload backhauls?
According to classical information theory, where the information is
considered as random raw bits, the answers to these questions are
negative. However, in practice, we are more concerned about delivering
content. By exploiting the fact that content is \textquotedblleft cachable\textquotedblright ,
we show that the answers to both questions can be positive using \textit{PHY
caching} at the BSs. Specifically, there are two fundamental benefits
associated with PHY caching in dense wireless networks. If the content
accessed by users exists simultaneously at the BS caches, the BSs
can engage in Co-MIMO and enjoy a large MIMO cooperation gain, as
illustrated in Fig. \ref{fig:system_model} for the red and green
data flows. This is referred to as \textit{cache-induced opportunistic
Co-MIMO}. If the content requested by a user is distributed in the
caches of the nearby BSs, this user can directly obtain the requested
content from the nearby BSs, which reduces the number of hops from
the source BSs to the destination user, as illustrated in Fig. \ref{fig:system_model}
for the purple data flow. This is referred to as \textit{cache-assisted
multihopping}.

We are interested in studying the fundamental linear capacity scaling
in the backhaul-limited C-DWN and how to exploit the above benefits
of PHY caching to achieve the linear capacity scaling. Some related
works are reviewed below.

\textbf{Capacity scaling law in wireless ad hoc networks:} The capacity
scaling law of wireless ad hoc networks was first studied by Gupta
and Kumar in the seminal paper \cite{GuptaKumar}, where they showed
that in a large wireless ad hoc networks with $N$ random located
nodes, the aggregate throughput of classical multihop communication
scheme scales at most as $\Theta\left(\sqrt{N}\right)$. After that,
a number of works \cite{Xie_liangliang_IT04_network_information_capacity,Jovicic_TIT2004_TCahoc,Kumar_TIT06_TCadhoc}
have studied the information theoretic capacity scaling law under
different channel models and traffic models. Specifically, it was
shown in \cite{Tse_IT07_CapscalingHMIMO} that under a physical channel
model with the path loss exponent $\alpha\in\left(2,3\right]$, the
total network capacity scales as $\Theta\left(N^{2-\alpha/2}\right)$,
which can approach the linear capacity scaling law $\Theta\left(N\right)$
as $\alpha$ approaches $2$. However, this capacity gain is at the
cost of increased system complexity due to the complicated \textit{hierarchical
MIMO cooperation} \cite{Tse_IT07_CapscalingHMIMO,Niesen_TIT09_CSadhoc}.
In \cite{Niesen_TIT09_CSadhoc}, the authors studied the capacity
scaling in ad hoc networks with arbitrary node placement. The capacity
regions of the ad hoc network with more complicated unicast or multicast
traffic model have been studied in \cite{Niesen_TIT10_CSadhoc}.

\textbf{Capacity scaling law in cellular networks:} Unlike ad hoc
networks, cellular networks consist of infrastructure gateways as
the sources of data traffics. When all BSs have wired payload backhauls,
the capacity scales linearly with the number of the BSs. However,
adding more \textit{backhaul-connected BSs} (i.e., BSs that have wired
backhauls) leads to high CAPEX and OPEX \cite{Paolini_mobiledatabackhaul}.
A more cost-effective way to enhance the capacity is to add relay
nodes to enable multihop communications between the users and BSs.
It was shown in \cite{Li_INFOCOM2011_capscalingMH} that multihop
cellular networks with $N$ relay nodes and $M$ BSs achieve higher
per-node throughput than pure cellular networks (with only $M$ BSs)
by a scaling factor of $\log_{2}N$. However, the total capacity order
$\Theta\left(M\log_{2}N\right)$ is still far from the linear capacity
$\Theta\left(N\right)$ for large $N$.

\textbf{Wireless Caching:} Recently, wireless caching has been proposed
as a cost-effective solution to handle the high traffic rate caused
by content delivery applications \cite{Goebbels_IJCS10_Smartcaching,Caire_INFOCOM12_femtocache}.
For example, \cite{Niesen_TIT12_FLcaching} proposed coded caching
schemes that can create coded multicast opportunities. A proactive
caching paradigm was proposed in \cite{Debbah_CoM14_proactivecache}
to exploit both the spatial and social structure of the wireless networks.
The fundamental tradeoff in wireless D2D caching networks have also
been studied in \cite{Caire_arxiv13_D2Dcaching,Caire_arxiv13_d2dcachingtradeoff,Altieri_arxiv14_d2dcaching,Jeon_ICC15_D2Dcaching}.
Furthermore, \cite{Gitzenis_TIT13_wirelesscache} studied the joint
optimization of cache content replication and routing in a \textit{regular
network} and identified the throughput scaling laws for various regimes.
Note that although both \cite{Gitzenis_TIT13_wirelesscache} and this
paper focus on the scaling law of cached wireless networks, they are
different in many aspects. First, the throughput scaling law in \cite{Gitzenis_TIT13_wirelesscache}
is obtained by assuming a specific caching and multihop transmission
scheme. On the other hand, the capacity scaling law studied in this
paper is an information theoretic scaling law which does not depend
on any specific caching or PHY transmission scheme. Second, the role
of caching is also different. In \cite{Gitzenis_TIT13_wirelesscache},
the main role of caching is to reduce the number of hops in multihop
transmissions. In this paper, the role of PHY caching includes both
the \textit{cache-assisted multihopping} and the \textit{cache-induced
MIMO cooperation}. Third, the network topologies are also different.
\cite{Gitzenis_TIT13_wirelesscache} considered a regular network
where all nodes are placed on a perfect grid and have homogeneous
traffic. This paper considers a general network with arbitrary node
placements and content requests. Finally, the key difference between
wireless and wired networks is that the performance of wireless networks
is fundamentally limited by the interference due to the broadcast
nature of wireless channel. However, this unique feature of wireless
networks and the associated PHY transmission scheme are not considered
in the analysis in \cite{Gitzenis_TIT13_wirelesscache}. In this paper,
we consider joint design of PHY caching and transmission schemes (e.g.,
the PHY transmission modes in our design depend on the cache mode
of the requested file) where interference plays a vital role (e.g.,
both frequency partitioning and cache-induced MIMO cooperation are
proposed to mitigate the interference). 

The above works on wireless caching do not consider cache-induced
MIMO cooperation among the BSs. The concept of cached-induced opportunistic
Co-MIMO was first introduced in \cite{Liu_TSP14_CacheRelay,Liu_TSP13_CacheIFN}.
The achievable throughput scaling laws of ad hoc networks with PHY
caching were also studied in \cite{Liu_TWC15arxiv_adhoccaching}.
However, the capacity scaling in the backhaul-limited C-DWN and the
corresponding order-optimal PHY caching and transmission schemes have
not been addressed in the literature. This paper provides solutions
to the following important questions associated with the backhaul-limited
C-DWN.
\begin{itemize}
\item \textbf{How much cache is needed to achieve the linear capacity scaling
in the backhaul-limited C-DWN? }We show that\textbf{ }the total network
capacity of the backhaul-limited C-DWN scales as $\Theta\left(N\right)$
when the BS cache size is larger than a threshold that depends on
the content popularity distribution. We quantify the minimum BS cache
size required to achieve the linear capacity scaling as a function
of the content popularity parameter. 
\item \textbf{What is the order-optimal capacity achieving scheme? }We propose
an order-optimal achievable scheme, which can exploit both the\textit{
}cache-induced opportunistic Co-MIMO and cache-assisted multihopping
to achieve the linear capacity scaling for the backhaul-limited C-DWN.
\item \textbf{What is the role of cache-induced Co-MIMO? }Exploiting cache-induced
Co-MIMO cannot change the throughput scaling law of the backhaul limited
C-DWN. However, it helps to mitigate the interference and increase
the spatial degrees of freedom (i.e., the number of data streams that
can be simultaneously transmitted) in the backhaul limited C-DWN.
As a result, a huge throughput gain can be achieved by exploiting
cache-induced Co-MIMO. In this paper, we derive closed-form expression
for this cache-induced MIMO cooperation gain and analyze the minimum
BS cache size needed to achieve a significant cache-induced MIMO cooperation
gain.
\end{itemize}

The rest of the paper is organized as follows. In Section \ref{sec:System-Model},
we introduce the system model. In Section \ref{sec:Uncached-Wireless-Network},
we give some preliminary results on capacity bound in backhaul-limited
dense wireless networks without cache. In Section \ref{sec:Capacity-Scaling-Law},
we present the main results on the linear capacity scaling law in
the backhaul-limited C-DWN. The order-optimal achievable schemes are
elaborated in Section \ref{sec:Order-wise-Optimal-Control}. The performance
analysis for the regular C-DWN is given in Section \ref{sec:Performance-Analysis}.
Section \ref{sec:Numerical-Results-and} provides some numerical results.
Section \ref{sec:Discussion} discusses some extensions and Section
\ref{sec:Conclusion} concludes.

\section{System Model\label{sec:System-Model}}

\subsection{Architecture of the Backhaul-Limited C-DWN}

\begin{figure}
\begin{centering}
\textsf{\includegraphics[clip,width=80mm]{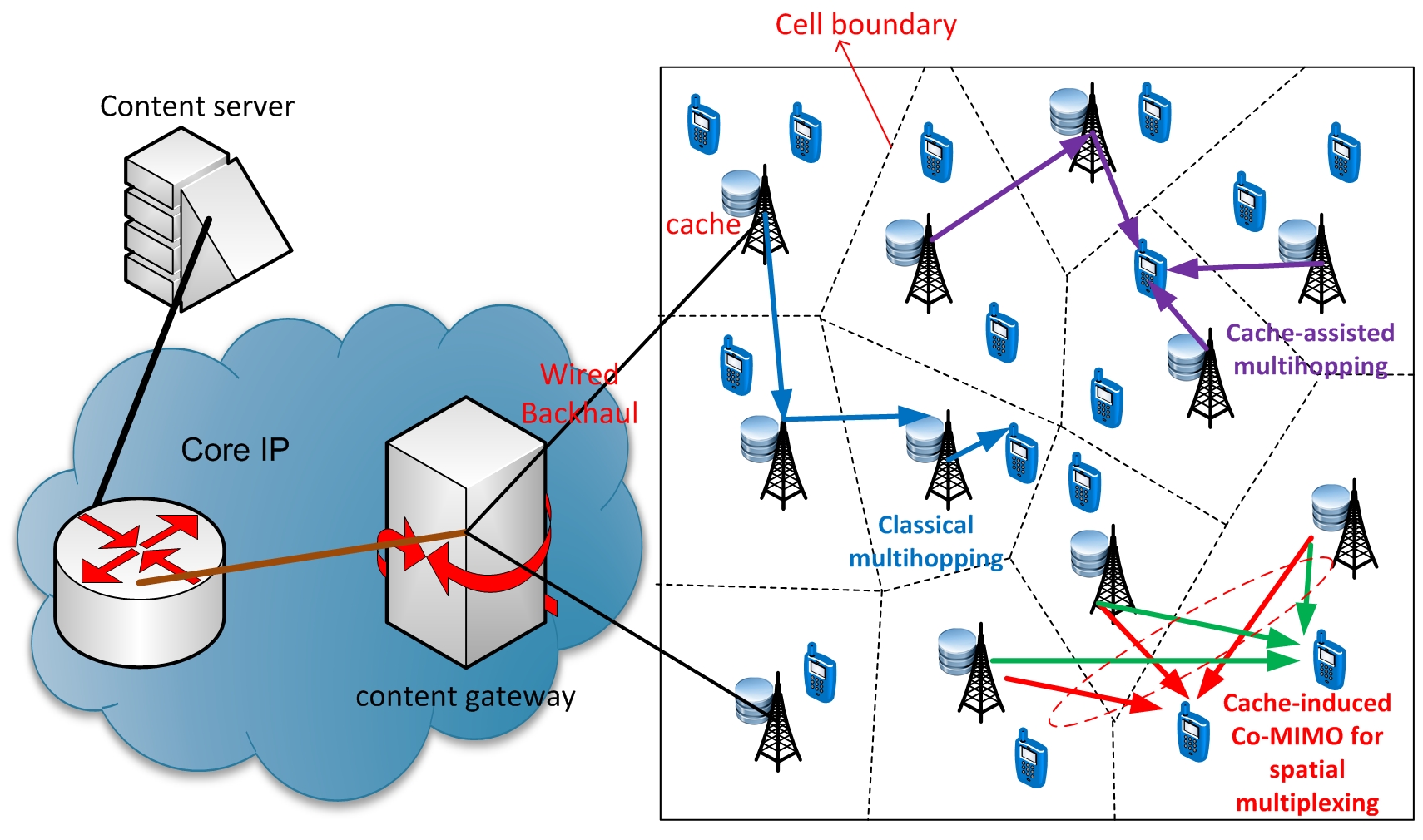}}
\par\end{centering}

\protect\caption{\label{fig:system_model}{\small{}Architecture of the backhaul-limited
C-DWN.}}
\end{figure}

Consider a backhaul-limited C-DWN with $N$ BSs and $K$ users placed
on a square of area $Nr_{0}^{2}$ as illustrated in Fig. \ref{fig:system_model}.
Each BS has an average transmit power budget of $P$ and a cache of
limited size $B_{C}$ bits. Only $N_{0}\ll N$ BSs have wired backhaul
connections with the core network. We divide the square into $N$
cells. The $n$-th cell is the set of all points which are closer
to the $n$-th BS. If a user $k$ lies in the $n$-th cell, user $k$
is said to be associated with BS $n$. Let $b_{k}$ denote the BS
associated with user $k$. We have the following assumption on the
BS and user placement.

\begin{assumption}[BS and User Placement]\label{asm:Place}The BS
and user placement satisfies the following conditions.
\begin{enumerate}
\item The distance between any two BSs is no less than $r_{\textrm{min}}$ 
\item The distance between a BS and a user is no less than $d_{\textrm{min}}$.
\item For any point in the $n$-th cell, the distance between this point
and the $n$-th BS is no more than $r_{\textrm{max}}$.
\item Each BS is associated with at most $k_{\textrm{max}}$ users, where
$k_{\textrm{max}}$ is constant.
\item The total numbers of users and BSs satisfy $K=\Theta\left(N\right)$.
\end{enumerate}
\end{assumption}

Assumption \ref{asm:Place}-1) means that the BS density at any local
area does not go to infinity, which is always satisfied in practice.
Assumption \ref{asm:Place}-2) means that the distance between any
user and its associated BS is bounded away from zero so that the receive
SNR at any user is bounded. Assumption \ref{asm:Place}-3) means that
the size of any cell is bounded so that the received SNR of any user
at any location is sufficiently large to establish a communication
link between this user and the network. This assumption is used to
ensure that there is no coverage hole in the network. Assumption \ref{asm:Place}-4)
means that the user density at each cell does not go to infinity.
Finally, Assumption \ref{asm:Place}-5) means that we consider a heavily
loaded system where there are active users in most of the cells (BSs).

A large portion of the traffic in future wireless networks will come
from content delivery applications where users obtain content (e.g.,
video) from the content server via the BSs and core network as illustrated
in Fig. \ref{fig:system_model}. There are $L$ content files on the
content server and the size of each file is $F$ bits%
\footnote{For clarity, we assume equal file size. The capacity scaling laws
in this paper also hold for the case when different files have different
sizes.%
}. For convenience, let $l_{k}$ denote the index of the file requested
by the $k$-th user and let $\vec{l}=\left\{ l_{1},...,l_{K}\right\} $
denote the user request profile (URP). Assume that each node independently
accesses the $l$-th content file with probability $p_{l}$, where
probability mass function $\mathbf{p}=\left[p_{1},...,p_{L}\right]$
represents the popularity of the content files. Without loss of generality,
suppose $p_{1}\geq p_{2}\geq...\geq p_{L}$. In the rest of the paper,
we focus on the non-trivial case when each BS does not have enough
cache to store all the $L$ files, i.e., $B_{C}<LF$.

\subsection{Cached and Uncached Dense Wireless Networks}

Suppose user $k$ is associated with a BS which has no wired backhaul.
Without PHY caching, user $k$ has to obtain the requested content
from the content server via the core network, wired backhaul, and
multihop wireless links, as illustrated in Fig. \ref{fig:system_model}
for the blue data flow. As a result, the user throughput is fundamentally
limited by the multihop wireless inter-BS communications. For convenience,
the backhaul-limited dense wireless networks without PHY caching is
called the \textit{backhaul-limited uncached dense wireless networks
(U-DWN)}. 

In this paper, we propose a new architecture called the \textit{cached
dense wireless network} \textit{(C-DWN)}, where the BSs can cache
some of the content from the content servers. If the content requested
by a user is in the caches of the nearby BSs, this user can directly
obtain the requested content from the nearby BSs via cache-induced
opportunistic Co-MIMO (as illustrated in Fig. \ref{fig:system_model}
for the red and green data flows) or cache-assisted multihopping (as
illustrated in Fig. \ref{fig:system_model} for the purple data flow),
depending on the \textit{cache state} of the nearby BSs. An interesting
question is that, what are the optimal capacity scaling and the corresponding
optimal achievable scheme in C-DWN. This question will be answered
in this paper.

\subsection{Channel Model}

We use similar channel model as in \cite{Tse_IT07_CapscalingHMIMO},
where the wireless link between any two nodes is a flat fading channel
with bandwidth $W$. The channel coefficient between BS $n$ and BS
$n^{'}$ at time $t$ is
\[
h_{n^{'},n}^{b}\left(t\right)=\sqrt{G^{b}}\left(r_{n^{'},n}^{b}\right)^{-\alpha/2}\exp\left(j\theta_{n^{'},n}^{b}\left(t\right)\right),
\]
where $r_{n^{'},n}^{b}$ is the distance between BS $n$ and BS $n^{'}$,
$\theta_{n^{'},n}^{b}\left(t\right)$ is the random phase at time
$t$, $G^{b}$ is some constant depending on the transmitter and receiver
antenna gains at the BSs, and $\alpha>2$ is the path loss exponent.
Similarly, the channel between BS $n$ and user $k$ at time $t$
is given by
\[
h_{k,n}^{d}\left(t\right)=\sqrt{G^{d}}\left(r_{k,n}^{d}\right)^{-\alpha/2}\exp\left(j\theta_{k,n}^{d}\left(t\right)\right),
\]
where $r_{k,n}^{d}$ is the distance between BS $n$ and user $k$,
$\theta_{k,n}^{d}\left(t\right)$ is the random phase, and $G^{d}$
is some constant. We assume that $\theta_{n^{'},n}^{b}\left(t\right)$
and $\theta_{k,n}^{d}\left(t\right)$ are i.i.d. (w.r.t. the node
index $n^{'},n,k$ and time index $t$) with uniform distribution
on $\left[0,2\pi\right]$. At each node, the received signal is corrupted
by a circularly symmetric Gaussian noise with spectral density $\eta_{0}$.

\subsection{Offline Cache Initialization\label{sub:Offline-Cache-Initialization}}

There are two phases during the operation of cached wireless networks:
the\textit{ cache initialization phase} and \textit{content delivery
phase}. In the cache initialization phase, each BS caches a portion
of $q_{l}F$ (possibly encoded) bits of the $l$-th content file ($\forall l$),
where $\mathbf{q}=\left[q_{1},...,q_{L}\right]^{T}$ (with $q_{l}\in\left[0,1\right]$
and $\sum_{l=1}^{L}q_{l}F\leq B_{C}$) are called \textit{cache content
replication vector}. The cache content replication vector $\mathbf{q}$
and the corresponding content stored at each BS is slowly adaptive
to the popularity of files according to some \textit{caching scheme}.
The cache initialization phase restarts whenever the popularity changes.
After each BS received the content determined by the caching scheme,
the content delivery phase starts. Let $t_{C}$ denote the interval
between two consecutive cache initialization phase. Since the popularity
of content files change very slowly (e.g., new movies are usually
posted on a weekly or monthly timescale), $t_{C}$ is large and thus
the cache update overhead in the cache initialization phase is usually
small compared to the PHY caching gain%
\footnote{This is a reasonable assumption widely used in the literature, see,
e.g., \cite{Gitzenis_TIT13_wirelesscache,Liu_TSP13_CacheIFN,Liu_TSP14_CacheRelay,Caire_arxiv13_d2dcachingtutorial}
and references there in.%
}. In this paper, we assume that $t_{C}$ is sufficiently large and
thus the cache update overhead can be ignored. We will focus on studying
the content delivery phase.

\section{Capacity Bound of Backhaul-Limited U-DWN\label{sec:Uncached-Wireless-Network}}

In this section, we give an information theoretic upper bound on the
aggregate throughput of the backhaul-limited U-DWN, which serves as
benchmarking to quantify the gain due to PHY caching later.

For convenience, let $\mathcal{B}_{P}$ denote the set of $N_{0}$\textit{
}backhaul-connected BSs. Let $\mathbf{H}_{b}=\left[h_{n^{'},n}^{b}\right]_{\forall n^{'}\in\left\{ 1,...,N\right\} \backslash\mathcal{B}_{P},n\in\mathcal{B}_{P}}\in\mathbb{C}^{\left(N-N_{0}\right)\times N_{0}}$
denote the composite channel between the $N_{0}$ backhaul-connected
BSs and all other BSs. Let $\mathbf{H}_{d}=\left[h_{k,n}^{d}\right]_{\forall k,n\in\mathcal{B}_{P}}\in\mathbb{C}^{K\times N_{0}}$
denote the composite channel between the $N_{0}$ backhaul-connected
BSs and all users. The following lemma is useful for deriving the
capacity upper bound of the backhaul-limited U-DWN.
\begin{lem}
\label{lem:cutB1}Let $\tilde{\mathbf{H}}=\left[\mathbf{H}_{b}^{T},\mathbf{H}_{d}^{T}\right]^{T}\in\mathbb{C}^{\left(N-N_{0}+K\right)\times N_{0}}$
denote the composite channel between the $N_{0}$ backhaul-connected
BSs and all other nodes. The aggregate throughput of the backhaul-limited
U-DWN is bounded above by
\begin{equation}
\tilde{T}\leq W\max_{\tilde{\mathbf{Q}}\left(\tilde{\mathbf{H}}\right)\succeq\mathbf{0},\textrm{E}\left[\textrm{Tr}\left(\tilde{\mathbf{Q}}\left(\tilde{\mathbf{H}}\right)\right)\right]\leq N_{0}P}\textrm{E}\left[\log\left|\mathbf{I}+\tilde{\mathbf{H}}\tilde{\mathbf{Q}}\left(\tilde{\mathbf{H}}\right)\tilde{\mathbf{H}}^{\dagger}\right|\right].\label{eq:Cutb-1}
\end{equation}
Moreover, we have $\textrm{Tr}\left(\tilde{\mathbf{H}}\tilde{\mathbf{H}}^{\dagger}\right)=\tilde{b}_{U}=\Theta\left(N_{0}\right),\forall\tilde{\mathbf{H}}$,
where
\[
\tilde{b}_{U}=\sum_{n\in\mathcal{B}_{P}}\sum_{n^{'}\notin\mathcal{B}_{P}}G^{b}\left(r_{n^{'},n}^{b}\right)^{-\alpha}+\sum_{n\in\mathcal{B}_{P}}\sum_{k=1}^{K}G^{d}\left(r_{k,n}^{d}\right)^{-\alpha}
\]

\end{lem}

Using Lemma \ref{lem:cutB1}, we can prove the following theorem.
\begin{thm}
[Capacity upper bound of backhaul-limited U-DWN]\label{thm:CUB-1}Define
a function 
\[
f\left(\xi,x\right)=\begin{cases}
\log\left(1+\xi x^{2}\right), & x\geq\sqrt{\frac{\zeta}{\xi}}\\
\frac{2\sqrt{\xi\zeta}}{1+\zeta}x, & x\in\left[0,\sqrt{\frac{\zeta}{\xi}}\right)
\end{cases},
\]
where $\xi=\frac{\tilde{b}_{U}}{N_{0}P}$, $\zeta=-1+e^{2+W_{0}\left(-2e^{-2}\right)}\approx3.9216$,
and $W_{0}\left(z\right)$ is the principal branch of the Lambert
W function. Then the aggregate throughput of the backhaul-limited
U-DWN is bounded above by
\[
\tilde{T}\leq N_{0}Wf\left(\xi,P\right)=\Theta\left(N_{0}\right).
\]

\end{thm}

Please refer to Appendix \ref{sub:Proof-of-CutB} for the proof of
Lemma \ref{lem:cutB1} and Theorem \ref{thm:CUB-1}.

\section{Capacity Scaling in Backhaul-Limited C-DWN\label{sec:Capacity-Scaling-Law}}

We first give an information theoretic upper bound on the aggregate
throughput. The following lemma is useful for deriving the capacity
upper bound.
\begin{lem}
\label{lem:cutB}Let $\mathbf{H}=\left[h_{k,n}^{d}\right]_{\forall k,n}\in\mathbb{C}^{K\times N}$
denote the composite channel between all BSs and all users. The aggregate
throughput of the backhaul-limited C-DWN is bounded by
\begin{equation}
T\leq W\max_{\mathbf{Q}\left(\mathbf{H}\right)\succeq\mathbf{0},\textrm{E}\left[\textrm{Tr}\left(\mathbf{Q}\left(\mathbf{H}\right)\right)\right]\leq NP}\textrm{E}\left[\log\left|\mathbf{I}+\mathbf{H}\mathbf{Q}\left(\mathbf{H}\right)\mathbf{H}^{\dagger}\right|\right].\label{eq:Cutb}
\end{equation}
Moreover, we have $\textrm{Tr}\left(\mathbf{H}\mathbf{H}^{\dagger}\right)=b_{U}=\Theta\left(N\right),\forall\mathbf{H}$,
where 
\[
b_{U}=\sum_{n=1}^{N}\sum_{k=1}^{K}G^{d}\left(r_{k,n}^{d}\right)^{-\alpha}.
\]

\end{lem}

Using Lemma \ref{lem:cutB}, we can prove the following result.
\begin{thm}
[Capacity upper bound of backhaul-limited C-DWN]\label{thm:CUB}The
aggregate throughput of the backhaul-limited C-DWN is bounded above
by
\[
T\leq NWf\left(\frac{b_{U}}{NP},P\right)=\Theta\left(N\right).
\]

\end{thm}

The proof of Lemma \ref{lem:cutB} and Theorem \ref{thm:CUB} is similar
to that of Lemma \ref{lem:cutB1} and Theorem \ref{thm:CUB-1} in
Appendix \ref{sub:Proof-of-CutB}. Next, we give the achievable throughput
scaling laws. Define $\tilde{L}=\frac{LF}{B_{C}}$ as the \textit{normalized
content size}. Let the notation $N,\tilde{L}\overset{\iota}{\rightarrow}\infty$
denote $N\rightarrow\infty$ and $\lim_{N\rightarrow\infty}\frac{\tilde{L}}{N}=\iota\in\left[0,\infty\right)$.
There are two classes of scaling laws depending on the normalized
content size $\tilde{L}$.
\begin{thm}
[Achievable throughput scaling laws with $\tilde{L}=\Theta\left(1\right)$]\label{thm:scalingL1}In
the backhaul-limited C-DWN, a per user throughput of $R=\Theta\left(1\right)$
is achievable when $N\rightarrow\infty$ and $\tilde{L}=\Theta\left(1\right)$.
As a result, an aggregate throughput of $KR=\Theta\left(N\right)$
is also achievable.
\end{thm}

When $\tilde{L}$ goes to infinity, the throughput scaling laws will
also depend on the content popularity distribution. We make the following
assumptions on the content popularity.

\begin{assumption}[Content popularity]\label{asm:Popdist}The content
popularity is modeled by the Zip distribution \cite{Yamakami_PDCAT06_Zipflaw}:
\begin{equation}
p_{l}=\frac{1}{Z_{\tau}\left(L\right)}l^{-\tau},l=1,...,L,\label{eq:zipp-1}
\end{equation}
where $\tau$ is the \textit{popularity skewness parameter}, and $Z_{\tau}\left(L\right)=\sum_{l=1}^{L}l^{-\tau}$
is a normalization factor.

\end{assumption} 
\begin{thm}
[Achievable throughput scaling laws with Large $\tilde{L}$]\label{thm:scalingLinf}In
the backhaul-limited C-DWN, a per user throughput of $R$ is achievable
when $N,\tilde{L}\overset{\iota}{\rightarrow}\infty$, where
\[
R=\begin{cases}
\Theta\left(1/\sqrt{\tilde{L}}\right), & if\:0\leq\tau<1\\
\min\left(\Theta\left(\log L/\sqrt{\tilde{L}}\right),\Theta\left(1\right)\right) & if\:\tau=1\\
\min\left(\Theta\left(L^{\tau-1}/\sqrt{\tilde{L}}\right),\Theta\left(1\right)\right) & if\:1<\tau<\frac{3}{2}\\
\min\left(\Theta\left(\sqrt{L}\log^{-3/2}L/\sqrt{\tilde{L}}\right),\Theta\left(1\right)\right) & if\:\tau=3/2\\
\Theta\left(1\right), & if\:\tau>3/2
\end{cases}
\]

\end{thm}

The proof of Theorem \ref{thm:scalingL1} and \ref{thm:scalingLinf}
relies on the construction of an achievable scheme that realizes the
promised scaling law. The details will be given in Section \ref{sec:Order-wise-Optimal-Control}.
Theorem \ref{thm:scalingLinf} and Theorem \ref{thm:CUB} together
establish the linear capacity scaling law in the backhaul-limited
C-DWN when the \textit{normalized cache size} $\tilde{B}_{C}=B_{C}/F$
is sufficiently large. Specifically, the following corollary from
Theorem \ref{thm:scalingL1} and \ref{thm:scalingLinf} quantifies
the minimum normalized cache size $\tilde{B}_{C}^{\textrm{min}}$
needed to achieve the linear capacity scaling law.
\begin{cor}
[Minimum cache size to achieve linear capacity scaling]\label{cor:Minimum-cache-size}When
$L=\Theta\left(1\right)$, the order of the minimum normalized cache
size needed to achieve the linear capacity scaling $R=\Theta\left(1\right)$
is $\tilde{B}_{C}^{\textrm{min}}=\Theta\left(1\right)$. When $N,\tilde{L}\overset{\iota}{\rightarrow}\infty$,
we have
\[
\tilde{B}_{C}^{\textrm{min}}=\begin{cases}
\Theta\left(L\right) & if\:0\leq\tau<1\\
\Theta\left(L/\log^{2}L\right) & if\:\tau=1\\
\Theta\left(L^{3-2\tau}\right) & if\:1<\tau<3/2\\
\Theta\left(\log^{3}L\right) & if\:\tau=3/2\\
\Theta\left(1\right) & if\:\tau>3/2
\end{cases}
\]

\end{cor}

\begin{figure}
\begin{centering}
\includegraphics[width=80mm]{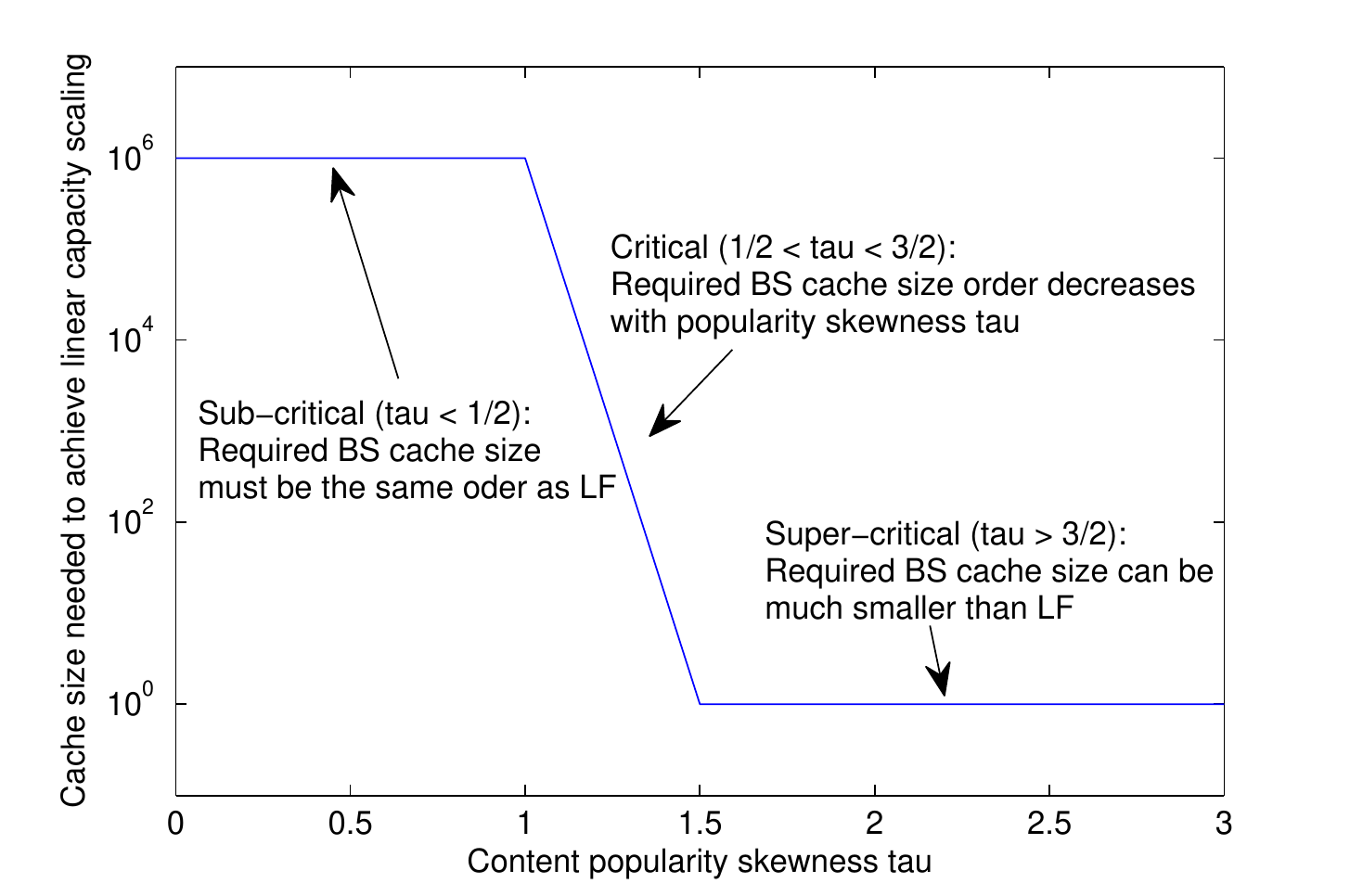}
\par\end{centering}

\protect\caption{\label{fig:sub-super-critical}Phase transition of minimum BS cache
size $B_{C}$ needed to achieve the linear capacity scaling for large
$L$.}
\end{figure}

For a larger $\tau$, the requests will concentrate more on a few
content files and thus a smaller BS cache is needed to achieve the
linear capacity scaling law. When $N,\tilde{L}\overset{\iota}{\rightarrow}\infty$,
there are two \textit{critical popularity skewness} points: $\tau=1$
and $\tau=3/2$, as illustrated in Fig. \ref{fig:sub-super-critical}.
For the super-critical case when $\tau>3/2$, the linear capacity
scaling laws can be achieved even when the cache size at each BS is
much smaller than the total content size (i.e., $B_{C}=\Theta\left(1\right)$
and $LF\rightarrow\infty$). The intuition behind this result is as
follows. When $\tau>3/2$, the user requests concentrate heavily on
the $\Theta\left(1\right)$ most popular files such that we can ignore
the impact of requesting the other $L-\Theta\left(1\right)$ files
on the throughput scaling law. In this case, we can achieve the linear
capacity scaling law by letting each BS cache a portion of $\frac{B_{C}}{F\Theta\left(1\right)}$
bits for each of the $\Theta\left(1\right)$ most popular files. This
is because under such caching scheme, whenever a user requests one
of the $\Theta\left(1\right)$ most popular files, the requested file
must exist in the nearest $\Theta\left(1\right)$ BSs and thus the
number of hops from the source BSs to the destination user is $\Theta\left(1\right)$.
As a result, we can achieve the linear capacity scaling since the
probability of requesting the other $L-\Theta\left(1\right)$ files
is negligible when $\tau>3/2$. On the other hand, for the sub-critical
case when $\tau<1$, the BS cache size $B_{C}$ must scale at the
same order as the content size $LF$ in order to achieve the linear
capacity scaling. The intuition behind this result is as follows.
When $\tau<1$, the popularity is very flat such that the order of
the minimum BS cache size $B_{C}$ needed to achieve the linear capacity
scaling is the same as that of the extremely case when the popularity
is completely flat (i.e., $p_{l}=\frac{1}{L},\forall l$). In this
case, the optimal caching scheme is uniformly caching (i.e., $q_{l}=\frac{B_{C}}{FL},\forall l$)
due to the symmetry of different files, and thus the requested file
of any user must exist in the nearest $\Theta\left(\frac{FL}{B_{C}}\right)$
BSs. Clearly, if we want to achieve the linear capacity scaling, we
must have $\frac{FL}{B_{C}}=\Theta\left(1\right)$, i.e., $B_{C}=\Theta\left(LF\right)$.

According to Theorem \ref{thm:scalingL1} and \ref{thm:scalingLinf},
we can achieve linear capacity scaling law in the backhaul-limited
C-DWN with only a fixed number of wired payload backhauls $N_{0}$
by a proper PHY caching scheme, when the cache size $B_{C}$ is sufficiently
large. Moreover, the required cache size to achieve the linear capacity
scaling law decreases with the popularity skewness $\tau$. These
results have fundamental impact on future wireless networks. Since
the cost of storage device is much lower than the cost of wired backhaul,
and the popularity skewness $\tau$ can be large for mobile applications
\cite{Yamakami_PDCAT06_Zipflaw}, the backhaul-limited C-DWN provides
a promising architecture for future wireless networks.

\section{Order-optimal Capacity Achieving Scheme\label{sec:Order-wise-Optimal-Control}}

In this section, we propose an achievable PHY caching and transmission
scheme (abbreviated as \textit{Scheme A}) which can achieve the throughput
scaling laws in Theorem \ref{thm:scalingL1} and \ref{thm:scalingLinf}.
There are two major components: the\textit{ two-mode maximum distance
separable} \textit{(MDS) coded caching} working in the cache initiation
phase and the \textit{cache-assisted PHY transmission} working in
the content delivery phase. The two-mode MDS-coded caching decides
how to cache the coded segments at each BS and the cache-assisted
PHY transmission contains two modes, namely, the \textit{cache-assisted
multihop transmission mode} and \textit{cache-induced Co-MIMO transmission
mode}. They are elaborated below.

\subsection{Two-mode MDS-coded Caching Scheme\label{sub:MDS-Cache-Encoding}}

The cache-assisted multihopping and cache-induced opportunistic Co-MIMO
have conflicting requirements on the caching scheme. For the former,
it is better to cache different content at different BSs. For the
later, it is better to cache the same content at nearby BSs to create
a MIMO cooperation opportunity. We propose a two-mode MDS-coded caching
scheme to strike a balance between them. There are two cache modes
depending on the popularity of the cached files. The Co-MIMO cache
mode is used for popular files to induce MIMO cooperation opportunity,
and the multihop cache mode is used for less popular files to achieve
cache-assisted multihopping gain. The details are summarized below. 

{\small{}S}\textbf{\small{}tep 1 (MDS Encoding and Cache Modes Determination):}{\small{}
At the content server, each file is divided into segments of size
$L_{S}$ bits and each segment is encoded using an ideal MDS rateless
code}%
\footnote{An MDS rateless code generates an arbitrarily long sequence of parity
bits from an information packet of $L_{S}$ bits, such that if the
decoder obtains any $L_{S}$ parity bits, it can recover the original
$L_{S}$ information bits \cite{Shokrollahi_TIT06_Raptorcode}.%
}{\small{}. If $q_{l}=1$, the cache mode for the $l$-th file is set
to be }\textit{\small{}Co-MIMO cache mode}{\small{}. In this case,
for each segment of the $l$-th file, the MDS encoder generates }\textit{\small{}one
Co-MIMO parity block}{\small{} of length $L_{S}$ bits. If $q_{l}<1$,
the cache mode for the $l$-th file is set to be }\textit{\small{}multihop
cache mode }{\small{}and the MDS encoder generates $N$ }\textit{\small{}multihop
parity blocks}{\small{} of length $q_{l}L_{S}$ for each segment of
the $l$-th file. Define $\Omega\left(\mathbf{q}\right)\triangleq\left\{ l:\: q_{l}<1\right\} $
as the set of files associated with multihop cache mode and $\overline{\Omega}\left(\mathbf{q}\right)\triangleq\left\{ l:\: q_{l}=1\right\} $
as the set of files associated with Co-MIMO cache mode.}{\small \par}

\textbf{\small{}Step 2 (Offline Cache Initialization):}{\small{} For
$l=1,..,L$, if $l\in\Omega\left(\mathbf{q}\right)$, then the cache
of BS $n$ is initialized with the $n$-th multihop parity block for
each segment of the $l$-th file. If $l\in\overline{\Omega}\left(\mathbf{q}\right)$,
then the caches of all the BSs are initialized with the Co-MIMO parity
block for each segment of the $l$-th file.}{\small \par}

We use an example to illustrate the above caching scheme with $2$
files. The first file is popular, and the Co-MIMO cache mode is used
($q_{1}=1$). The second file is less popular, and the multihop cache
mode is used ($q_{2}=0.5$). The size of each file is $3$Mbits and
the segment size is $L_{S}=1$Mbits. There are two BSs and the cache
size $B_{C}$ is $4.5$Mbits. For file 1, the content server only
generates a single Co-MIMO parity block of length 1Mbits from each
segment (there are totally 3 Co-MIMO parity blocks). BS 1 and BS 2
cache the same Co-MIMO parity block of every segment of file 1. For
file 2, the content server generates two multihop parity blocks of
length 0.5Mbits from each segment (there are totally 6 multihop parity
blocks). BS 1 (2) caches the first (second) multihop parity block
of every segment of file 2. 

The achievable throughput depends on the choice of the cache content
replication vector $\mathbf{q}$. In Theorem \ref{thm:CLB}, we will
give an \textit{order-optimal} cache content replication vector $\mathbf{q}$
to maximize the order of the per user throughput.

\subsection{PHY Mode Determination and Frequency Partitioning}

The transmission mode for a requested file is determined by the cache
mode. Specifically, a requested file $l\in\Omega\left(\mathbf{q}\right)$
($l\in\overline{\Omega}\left(\mathbf{q}\right)$) associated with
the multihop cache mode (Co-MIMO cache mode) is delivered to the destination
user using the cache-assisted multihop transmission (cache-induced
Co-MIMO). To transmit a file $l\in\Omega\left(\mathbf{q}\right)$
to a user, the associated BS needs to first collect the requested
parity bits from a set of source BSs (determined by \textit{source
BS set selection}) via \textit{wireless inter-BS transmission} and
then send them to this user via \textit{downlink access transmission},
as illustrated in Fig. \ref{fig:Multihop_Tx}. Correspondingly, the
system bandwidth $W$ is divided into three bands: the \textit{inter-BS
band} with size $W_{b}$ for wireless inter-BS transmission, the\textit{
downlink access band} with size $W_{d}$ for downlink access transmission,
and the\textit{ Co-MIMO band} with size $W_{c}$ for Co-MIMO transmission,
where $W_{b}+W_{d}+W_{c}=W$. As a result, these three transmissions
can occur simultaneously without causing interference to each other.

\subsection{Cache-assisted Multihop Transmission}

\subsubsection{Source BS Set Selection}

For a user $k$ requesting a file $l\in\Omega\left(\mathbf{q}\right)$,
the associated BS needs to select a set of source BSs to download
the requested file segment. Intuitively, the associated BS should
select the \textit{source BS set} so as to reduce the number of hops
and to balance the BS loads. We propose the following scheme to achieve
this. 

\textbf{\small{}Step 1 (Source BS Set Selection):}{\small{} The associated
BS chooses the nearest BSs (including the associated BS) which have
a total number of $L_{S}$ parity bits of the requested file segment
as the source BSs. Specifically, let $r_{k}^{*}=\min\: r,\:\textrm{s.t. }\left|\left\{ n:\: r_{b_{k},n}\leq r\right\} \right|\geq\left\lceil 1/q_{l}\right\rceil $.
Then the set of source BSs for user $k$ is given by $\mathcal{B}_{k}=\left\{ n:\: r_{b_{k},n}\leq r_{k}^{*}\right\} $.}{\small \par}

\textbf{\small{}Step 2 (Load Partitioning):}{\small{} The associated
BS determines the load partition among the source BSs as follows.
For convenience, define $\overline{\mathcal{B}}_{k}=\left\{ n:\: r_{b_{k},n}<r_{k}^{*}\right\} $.
Note that $\left|\overline{\mathcal{B}}_{k}\right|\leq\left\lceil 1/q_{l}\right\rceil -1$.
For each requested file segment, user $k$ obtains $q_{l}L_{S}$ parity
bits from each BS in $\overline{\mathcal{B}}_{k}$ and $\frac{\left(1-\left|\overline{\mathcal{B}}_{k}\right|q_{l}\right)L_{S}}{\left|\mathcal{B}_{k}\right|-\left|\overline{\mathcal{B}}_{k}\right|}$
parity bits from each BS in $\mathcal{B}_{k}\backslash\overline{\mathcal{B}}_{k}$.}{\small \par}

The following lemma gives an upper bound for $r_{k}^{*}$.
\begin{lem}
\label{lem:SBSbound}For a user $k$ requesting a file $l\in\Omega\left(\mathbf{q}\right)$
associated with the multihop cache mode, we have 
\[
r_{k}^{*}\leq\left(2\sqrt{\left\lceil 1/q_{l}\right\rceil -1}+1\right)r_{\textrm{max}}.
\]

\end{lem}

Please refer to Appendix \ref{sub:Proof-of-LemmaCM} for the proof.

\subsubsection{Wireless Inter-BS Transmission\label{sub:Wireless-Inter-BS-Transmission}}

Frequency reuse is used to control the interference between the BSs
on the inter-BS band. The inter-BS bandwidth $W_{b}$ is uniformly
divided into $M_{b}$ subbands and each BS is allocated with one subband
such that the following condition is satisfied. 

\begin{condition}\label{cond:FR}Any two BSs with distance no more
than $r_{I}^{b}$ is allocated with different subbands, where $r_{I}^{b}>2r_{\textrm{max}}$
is a system parameter. 

\end{condition}

The following lemma gives the number of subbands that is required
to satisfy the above condition. 
\begin{lem}
\label{lem:FRboundM}There exists a frequency reuse scheme which has
$M_{b}\leq\left(\frac{2r_{I}^{b}}{r_{\textrm{min}}}+1\right)^{2}+1$
subbands and satisfies Condition \ref{cond:FR}.
\end{lem}

Please refer to Appendix \ref{sub:Proof-of-LemmaCM} for the proof.

\begin{figure}
\begin{centering}
\textsf{\includegraphics[clip,width=80mm]{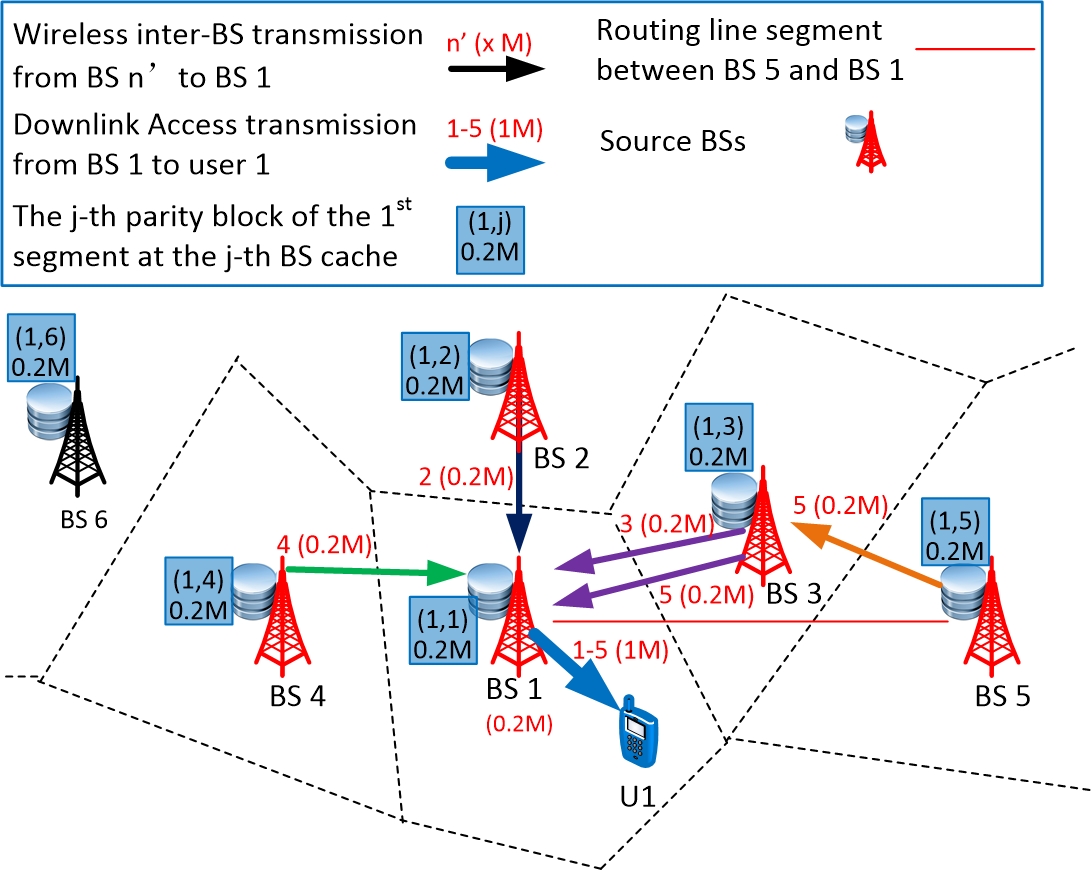}}
\par\end{centering}

\protect\caption{\label{fig:Multihop_Tx}{\small{}An illustration of cache-assisted
multihop transmission, where user $1$ requests the first segment
(with size $L_{S}=1$Mbits) of a file associated with multihop caching
mode. The requested file segment is encoded into parity blocks of
0.2Mbits at the content server and each BS caches one parity block
(0.2M parity bits). }}
\end{figure}

We now elaborate the wireless inter-BS transmission scheme for a user
$k$ requesting a file $l\in\Omega\left(\mathbf{q}\right)$. For any
source BS $n^{'}$other than the associated BS $n=b_{k}$, we draw
a \textit{routing line segment} $\mathcal{L}_{n^{'},n}$ between BS
$n^{'}$ and BS $n$. This routing line segment intersects several
cells. Then the parity bits requested by the user are relayed from
BS $n^{'}$ to the BS $n$ in a sequence of hops. In each hop, the
parity bits are transferred from one cell (BS) to another in the order
in which they intersect the routing line segment, as illustrated in
Fig. \ref{fig:Multihop_Tx}.

For convenience, we call $\left\{ \mathcal{L}_{n^{'},b_{k}},\forall n^{'}\in\mathcal{B}_{k}\backslash\left\{ b_{k}\right\} \right\} $
the set of routing line segments of user $k$. The following lemma
is useful when deriving the lower bound for the achievable rates.
\begin{lem}
\label{lem:bounJ}For any $n$, the average number of users whose
routing line segments intersect cell $n$ is upper bounded by 
\[
J\triangleq\sum_{l=1}^{L}p_{l}\left(\frac{\left(4\sqrt{\left\lceil 1/q_{l}\right\rceil -1}+2\right)r_{\textrm{max}}}{r_{\textrm{min}}}+1\right)^{2}k_{\textrm{max}}.
\]

\end{lem}

Please refer to Appendix \ref{sub:Proof-of-LemmaCM} for the proof.

\subsubsection{Downlink Access Transmission for Cache-assisted Multihop\label{sub:Downlink-Access-Transmission}}

For a user $k$ requesting a file $l\in\Omega\left(\mathbf{q}\right)$,
the wireless inter-BS transmission scheme ensures that all the segments
(parity bits) requested by user $k$ are available at the associated
BS $b_{k}$. Then the associated BS $b_{k}$ sends the requested parity
bits to user $k$ using downlink access transmission. To control the
inter-cell interference, the downlink access bandwidth $W_{d}$ is
uniformly divided into $M_{d}\leq\left(\frac{2r_{I}^{d}}{r_{\textrm{min}}}+1\right)^{2}+1$
subbands and each BS is allocated with one subband such that any two
BSs with distance no more than $r_{I}^{d}$ is allocated with different
subbands, where $r_{I}^{d}>r_{\textrm{max}}$ is a system parameter.
Within each cell, a simple TDMA scheme is used to mitigate the multi-user
interference in the downlink access transmission, where at each time
slot, only one user is scheduled for transmission in a round robin
fashion. 

The overall cache-assisted multihop transmission is summarized in
Fig. \ref{fig:Multihop_Tx}. The cache-assisted multihop transmission
contains 3 steps. 

\textbf{\small{}Step 1 (source BS set selection):}{\small{} The associated
BS (BS 1) chooses the nearest BSs (including BS 1) which have a total
number of 1M parity bits of the requested file segment as the source
BS set (which is $\left\{ 1,...,5\right\} $). The load partition
among the source BSs is illustrated in red above each data flow (colored
arrow). }\textbf{\small{}Step 2 (wireless inter-BS transmission):}{\small{}
The wireless inter-BS transmission between a source BS, say BS 5,
and the associated BS is as follows. We first draw a routing line
segment between BS 5 and BS 1 (red line). The routing line segment
intersects cell 3 and cell 1. Then BS 5 sends the parity bits requested
by user $1$ to the associated BS (BS 1) via wireless inter-BS transmission
over the route BS 5$\rightarrow$BS 3$\rightarrow$BS 1. }\textbf{\small{}Step
3 (downlink access transmission):}{\small{} BS 1 transmits the collected
1M parity bits (0.2Mbits from the local cache and 0.8Mbits from the
other source BSs) of the requested file segment to user $1$ via downlink
access transmission. Note that the arrows with different colors represent
the transmissions on different subbands for interference control.}{\small \par}

\subsection{Cache-induced Co-MIMO Transmission\label{sub:Cache-induced-Co-MIMO-Transmissi}}

For a user requesting a file $l\in\overline{\Omega}\left(\mathbf{q}\right)$,
the requested segment exists at all BS caches. As a result, the BSs
can employ Co-MIMO to improve the PHY performance. The cache-induced
Co-MIMO scheme contains the following three steps.

\textbf{\small{}Step 1 (BS clustering):}{\small{} At each time slot,
the whole area is partitioned into squares of size $N_{c}r_{0}^{2}$,
where $N_{c}$ determines the BS cluster size. Then the BSs in the
same square forms a cluster. }{\small \par}

\textbf{\small{}Step 2 (User scheduling in each cluster):}{\small{}
Without loss of generality, we consider the $j$-th cluster. Let $\mathcal{G}_{j}^{B}$
denote the set of BSs in the $j$-th cluster and let $\mathcal{G}_{j}^{K}\triangleq\left\{ k:\: b_{k}\in\mathcal{G}_{j}^{B},l_{k}\in\overline{\Omega}\left(\mathbf{q}\right)\right\} $
denote the set of users which are associated with the BSs in $\mathcal{G}_{j}^{B}$.
At each time slot, $K_{j}^{S}=\min\left(\left|\mathcal{G}_{j}^{K}\right|,\left|\mathcal{G}_{j}^{B}\right|\right)$
users in $\mathcal{G}_{j}^{K}$ are scheduled for transmission in
a round robin fashion.}{\small \par}

\textbf{\small{}Step 3 (Co-MIMO transmission in each cluster):}{\small{}
In the $j$-th cluster, the $\left|\mathcal{G}_{j}^{B}\right|$ BSs
employ Co-MIMO to jointly transmit some parity bits to the scheduled
$K_{j}^{S}$ users.} 

Finally, we adopt uniform power allocation where the power allocated
to each subband is proportional to the bandwidth of the subband. For
example, at each BS, the power allocated on each subband of the wireless
inter-BS transmission is $\frac{W_{b}P}{M_{b}W^{'}}$, the power allocated
on each subband of the downlink access transmission is $\frac{W_{d}P}{M_{d}W^{'}}$,
and the power allocated on the Co-MIMO band is $\frac{W_{c}P}{W^{'}}$,
where $W^{'}=\frac{W_{b}}{M_{b}}+\frac{W_{d}}{M_{d}}+W_{c}$. The
total transmit power of a BS is given by $P$.

\subsection{Order Optimality Analysis}

The order optimality of Scheme A is summarized in the following theorem.
\begin{thm}
[Order optimality of Scheme A]\label{thm:CLB}In the backhaul-limited
C-DWN, a per user throughput of 
\begin{equation}
R=\Theta\left(\frac{W}{\sum_{l=1}^{L}p_{l}\sqrt{\frac{1}{q_{l}}}}\right)\label{eq:Rorder}
\end{equation}
 can be achieved by Scheme A. Moreover, the order-optimal cache content
replication vectors $\mathbf{q}^{*}$ and the corresponding per user
throughput orders for different cases are given below.
\begin{enumerate}
\item When the normalized content size $\tilde{L}=\Theta\left(1\right)$
and $N\rightarrow\infty$, an order-optimal cache content replication
vector is given by $q_{l}^{*}=\frac{B_{C}}{LF},l=1,...,L$, and the
corresponding per user throughput order is $\Theta\left(1\right)$.
\item When $N,\tilde{L}\overset{\iota}{\rightarrow}\infty$, an order-optimal
cache content replication vector is given by 
\begin{equation}
q_{l}^{*}=\min\left(\frac{B_{C}}{F}\frac{p_{l}^{2/3}}{\sum_{l=1}^{L}p_{l}^{2/3}},1\right),l=1,...,L,\label{eq:orderq}
\end{equation}
and the corresponding per user throughput order is the same as that
in Theorem \ref{thm:scalingLinf}.
\end{enumerate}
\end{thm}

The factor $\sum_{l=1}^{L}p_{l}\sqrt{\frac{1}{q_{l}}}$ in (\ref{eq:Rorder})
is due to the wireless inter-BS transmission and it determines the
order of per user throughput. According to Lemma \ref{lem:bounJ},
the traffic to be relayed by a BS due to wireless inter-BS transmission
is upper bounded by $JR$, where $J=\Theta\left(\sum_{l=1}^{L}p_{l}\sqrt{\frac{1}{q_{l}}}\right)$.
Since the capacity of a BS on the inter-BS band is $\Theta\left(1\right)$,
we must have $R\leq\Theta\left(1/J\right)$. Hence, the capacity order
of the backhaul-limited C-DWN is mainly limited by the wireless inter-BS
transmission. Please refer to Appendix \ref{sub:Proof-of-TheoremCLB}
for the detailed proof of Theorem \ref{thm:CLB}.

\section{What is the Role of Cache-induced Co-MIMO?\label{sec:Performance-Analysis}}

Since the cache-induced Co-MIMO cannot improve scaling laws, an important
question is that, what is the role of cache-induced Co-MIMO and is
it worthwhile to\textbf{ }exploit cache-induced Co-MIMO? In this section,
we are going to answer this question by comparing Scheme A with a
baseline scheme called \textit{Scheme B}, which exploits purely the
cached-assisted multihopping benefit in C-DWN. In Scheme B, there
is no cache-induced Co-MIMO transmission as depicted in Section \ref{sub:Cache-induced-Co-MIMO-Transmissi}.
As a result, we have $W_{c}=0$ and $W_{b}+W_{d}=W$. The PHY transmission
scheme for requesting a file $l\in\Omega\left(\mathbf{q}\right)$
is based on the cache-assisted multihop transmission summarized in
Fig. \ref{fig:Multihop_Tx}. On the other hand, a requested file $l\in\overline{\Omega}\left(\mathbf{q}\right)$
is delivered from the associated BS to the destination user using
the downlink access transmission described in \ref{sub:Downlink-Access-Transmission}
since all the segments of a file $l\in\overline{\Omega}\left(\mathbf{q}\right)$
exist in the caches of all BSs. 

We will first analyze the per BS throughput of Scheme A and B in the
regular C-DWN defined below.
\begin{defn}
[Regular C-DWN]In a regular C-DWN, the BSs are placed on a grid
as illustrated in Fig. \ref{fig:grid-Tx}. The distance between the
adjacent BSs is $r_{0}$. Each BS has four users and they are placed
on the grid line around the BS. The distance between a user and the
associated BS is $d_{0}<r_{0}/2$.
\end{defn}

Then we will quantify the cache-induced MIMO cooperation gain, which
is defined as the per BS throughput gap between Scheme A and Scheme
B. In the following analysis, we let $N\rightarrow\infty$ to get
rid of the boundary effect. We also assume symmetric traffic model
where all users have the same throughput requirement $R$.

\subsection{Closed-form Bounds for Per BS Throughput}

\begin{figure}
\begin{centering}
\textsf{\includegraphics[clip,width=80mm]{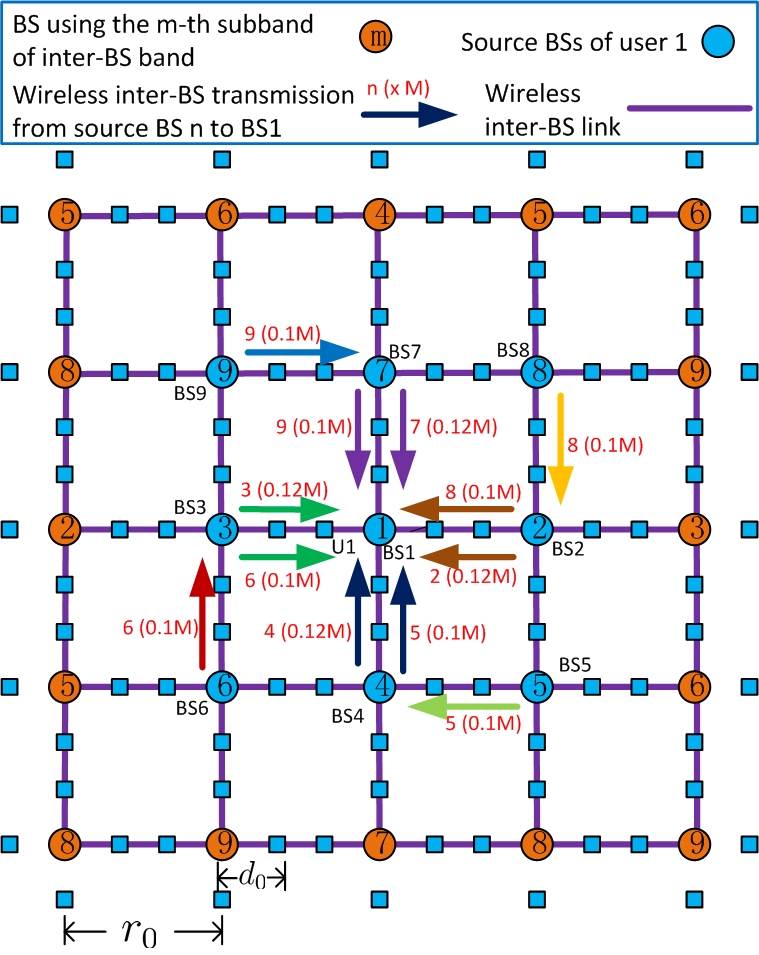}}
\par\end{centering}

\protect\caption{\label{fig:grid-Tx}{\small{}An illustration of cache-assisted multihop
transmission in the regular C-DWN. User $1$ requests the $i$-th
segment (with size $L_{S}=1$Mbits) of file $l$ associated with multihop
caching mode. The requested file segment is encoded into parity blocks
of 0.12Mbits at the content server and each BS caches one parity block
(0.12M parity bits). The set of source BSs for user $1$ is $\left\{ 1,...,9\right\} $.
The wireless inter-BS transmissions between the source BSs and the
associated BS (BS1) are illustrated with colored arrows, where different
colors represents different subbands. After collecting $1$M parity
bits of the $i$-th segment from the source BSs, BS 1 transmits them
to user 1 via downlink access transmission.}}
\end{figure}

It is highly non-trivial to derive the exact expression for the per
BS throughput especially for Scheme A with complicated cache-induced
Co-MIMO transmission. In this section, we derive closed-form bounds
for per BS throughput which are asymptotically tight at high SNR.
The common parameters in Scheme A and B are set as: $r_{I}^{b}=2.5r_{0}$
and $r_{I}^{d}=1.5r_{0}$. As a result, the inter-BS bandwidth $W_{b}$
is divided into $M_{b}=9$ subbands and the distance between a BS
and its nearest interfering BS is $3r_{0}$ on the inter-BS band,
as illustrated in Fig. \ref{fig:grid-Tx}. On the other hand, $M_{d}=4$
and it can be verified that the distance between a BS and its nearest
interfering BS is $2r_{0}$ on the downlink access band. In Fig. \ref{fig:grid-Tx},
we illustrate the cache-assisted multihop transmission for a regular
C-DWN. According to the wireless inter-BS transmission scheme in Section
\ref{sub:Wireless-Inter-BS-Transmission}, only the adjacent BSs can
communicate with each other and the wireless link between two adjacent
BSs is called a \textit{wireless} \textit{inter-BS link} as illustrated
in Fig. \ref{fig:grid-Tx}. 

First, we derive the average rate of each wireless inter-BS link and
the per user downlink access transmission rate in the cache-assisted
multihop transmission. 
\begin{lem}
\label{lem:RbA}The average rate of each wireless inter-BS link is
$W_{b}R_{b}\left(W_{b},W_{c}\right)$, where 
\[
R_{b}\left(W_{b},W_{c}\right)=\frac{1}{36}\log\left(1+\frac{PG^{b}r_{0}^{-\alpha}}{W^{'}\eta_{0}+PG^{b}I_{R}\left(1,3\right)}\right),
\]
$W^{'}=\left(9W-5W_{b}+27W_{c}\right)/36$, and 

\begin{eqnarray*}
 &  & I_{R}\left(x,y\right)\\
 & = & \sum_{i=1}^{\infty}\frac{1}{r_{0}^{\alpha}}\left[\sum_{j=1}^{\infty}2\left(\left(yi+x\right)^{2}+y^{2}j^{2}\right)^{-\frac{\alpha}{2}}+\left|yi+x\right|^{-\alpha}\right]\\
 & + & \sum_{i=1}^{\infty}\frac{1}{r_{0}^{\alpha}}\left[\sum_{j=1}^{\infty}2\left(\left(yi-x\right)^{2}+y^{2}j^{2}\right)^{-\frac{\alpha}{2}}+\left|yi-x\right|^{-\alpha}\right]\\
 & + & \sum_{i=1}^{\infty}\frac{2}{r_{0}^{\alpha}}\left(y^{2}i^{2}+x^{2}\right)^{-\frac{\alpha}{2}}.
\end{eqnarray*}
The per user average downlink access transmission rate is given by
$W_{d}R_{d}\left(W_{b},W_{c}\right)$, where
\[
R_{d}\left(W_{b},W_{c}\right)=\frac{1}{16}\log\left(1+\frac{PG^{d}d_{0}^{-\alpha}}{W^{'}\eta_{0}+PG^{d}I_{R}\left(\frac{d_{0}}{r_{0}},2\right)}\right).
\]

\end{lem}

Please refer to Appendix \ref{sub:Proof-of-LemmaRbARdA} for the proof
of Lemma \ref{lem:RbA}.

Clearly, $R_{b}$ is bounded as $R_{b}^{U}\geq R_{b}\left(W_{b},W_{c}\right)\geq R_{b}^{L}$,
where 
\[
R_{b}^{U}=\frac{1}{36}\log\left(1+\frac{9PG^{b}r_{0}^{-\alpha}}{W\eta_{0}+9PG^{b}I_{R}\left(1,3\right)}\right),
\]
\begin{eqnarray*}
R_{b}^{L} & = & \frac{1}{36}\log\left(1+\frac{PG^{b}r_{0}^{-\alpha}}{W\eta_{0}+PG^{b}I_{R}\left(1,3\right)}\right),
\end{eqnarray*}
and $R_{d}$ is bounded as $R_{d}^{U}\geq R_{d}\left(W_{b},W_{c}\right)\geq R_{d}^{L}$,
where 
\[
R_{d}^{U}=\frac{1}{16}\log\left(1+\frac{9PG^{d}d_{0}^{-\alpha}}{W\eta_{0}+9PG^{d}I_{R}\left(\frac{d_{0}}{r_{0}},2\right)}\right),
\]
\begin{eqnarray*}
R_{d}^{L} & = & \frac{1}{16}\log\left(1+\frac{PG^{d}d_{0}^{-\alpha}}{W\eta_{0}+PG^{d}I_{R}\left(\frac{d_{0}}{r_{0}},2\right)}\right).
\end{eqnarray*}
On the other hand, the average rate of a user on the Co-MIMO band
has no closed-form expression. The following lemma gives closed-form
bounds for the average Co-MIMO transmission rate.
\begin{lem}
[Average rate bounds for cache-induced Co-MIMO]\label{lem:LoCoMIMObound}Let
$G_{C}=I_{R}\left(\frac{d_{0}}{r_{0}},1\right)+d_{0}^{-\alpha}$,
$\rho=\frac{1}{2}\left(1-\frac{d_{0}^{-\alpha}}{G_{C}}\right)^{2}$,
$R_{c}^{U}=\frac{1}{4}\log\left(1+\frac{9PG^{d}G_{C}}{W\eta_{0}}\right)$
and $R_{c}^{L}=\frac{\rho}{4}\log\left(1+\frac{PG^{d}d_{0}^{-\alpha}}{W\eta_{0}}\right)$.
The average rate of a user on the Co-MIMO band is $W_{c}R_{c}\left(W_{b},W_{c}\right)$
, where $R_{c}\left(W_{b},W_{c}\right)$ is bounded as 
\[
R_{c}^{U}\geq R_{c}\left(W_{b},W_{c}\right)\geq R_{c}^{L}+O\left(PN_{c}^{-\frac{\alpha-2}{2\left(\alpha-1\right)}}\right).
\]

\end{lem}

Please refer to Appendix \ref{sub:Proof-of-TheoremCoMIMO} for the
proof. Finally, using Lemma \ref{lem:RbA} and \ref{lem:LoCoMIMObound},
the per BS throughput of Scheme A is bounded in the following theorem.
\begin{thm}
[Per BS throughput bounds of Scheme A]\label{thm:PBTB}The per BS
throughput $\Gamma_{A}\left(\mathbf{q}\right)$ of Scheme A is bounded
as $\Gamma_{A}^{U}\left(\mathbf{q}\right)\geq\Gamma_{A}\left(\mathbf{q}\right)\geq\Gamma_{A}^{L}\left(\mathbf{q}\right)+O\left(PN_{c}^{-\frac{\alpha-2}{2\left(\alpha-1\right)}}\right)$
with 
\begin{equation}
\Gamma_{A}^{a}\left(\mathbf{q}\right)=\frac{4R_{b}^{a}R_{c}^{a}R_{d}^{a}W}{Q_{\mathbf{q}}^{b}R_{c}^{a}R_{d}^{a}+Q_{\mathbf{q}}^{c}R_{b}^{a}R_{d}^{a}+Q_{\mathbf{q}}^{d}R_{b}^{a}R_{c}^{a}},\label{eq:Taqgen}
\end{equation}
for $a\in\left\{ L,U\right\} $, where $Q_{\mathbf{q}}^{b}=\sum_{l\in\Omega\left(\mathbf{q}\right)}p_{l}\psi\left(q_{l}\right)$,
$Q_{\mathbf{q}}^{c}=\sum_{l\in\overline{\Omega}\left(\mathbf{q}\right)}p_{l}$,
$Q_{\mathbf{q}}^{d}=\sum_{l\in\Omega\left(\mathbf{q}\right)}p_{l}$,
and 
\begin{eqnarray}
\psi\left(q_{l}\right) & = & \phi\left(q_{l}\right)\left(1-q_{l}\right)-\frac{2}{3}\left(\phi^{3}\left(q_{l}\right)-\phi\left(q_{l}\right)\right)q_{l},\label{eq:RbA}\\
\phi\left(q_{l}\right) & = & \left\lceil \frac{-1+\sqrt{\frac{2}{q_{l}}-1}}{2}\right\rceil .\nonumber 
\end{eqnarray}
Finally, as $P,N_{c}\rightarrow\infty$ such that $PN_{c}^{-\frac{\alpha-2}{2\left(\alpha-1\right)}}\rightarrow0$,
we have 
\begin{equation}
\Gamma_{A}\left(\mathbf{q}\right)\rightarrow\frac{W\tilde{R}_{b}\tilde{R}_{d}}{4Q_{\mathbf{q}}^{d}\tilde{R}_{b}+9Q_{\mathbf{q}}^{b}\tilde{R}_{d}},\label{eq:TAq}
\end{equation}
where $\tilde{R}_{b}=\log\left(1+\frac{r_{0}^{-\alpha}}{I_{R}\left(1,3\right)}\right)$
and $\tilde{R}_{d}=\log\left(1+\frac{d_{0}^{-\alpha}}{I_{R}\left(\frac{d_{0}}{r_{0}},2\right)}\right)$.
\end{thm}

Please refer to Appendix \ref{sub:Proof-of-TheoremPBTB} for the proof.
The physical meaning of the terms $\phi\left(q_{l}\right)$ and $\psi\left(q_{l}\right)$
in Theorem \ref{thm:PBTB} can be interpreted as follows. As can be
seen in Fig. \ref{fig:grid-Tx}, for each BS, the number of BSs with
the nearest distance ($r_{0}$) from the associated BS is $4$, that
with the second nearest distance ($\sqrt{2}r_{0}$) is $8$, and that
with the $m$-th nearest distance is $4m$. Let $\mathcal{B}_{k,m}$
denote the set of BSs with the $m$-th nearest distance from the associated
BS. Suppose user $k$ requests file $l$. Then $\phi\left(q_{l}\right)$
is the maximum number of hops between the associate BS and the source
BSs in $\mathcal{B}_{k}$ (e.g., in Fig. \ref{fig:grid-Tx}, the maximum
number of hops between the associate BS of user $1$ and its source
BSs is $\phi\left(0.12\right)=2$). Moreover, it can be shown that
$\sum_{l\in\Omega\left(\mathbf{q}\right)}p_{l}\psi\left(q_{l}\right)R$
is the average traffic rate on the inter-BS band induced by a single
user. 

Following similar analysis, it can be shown that the per BS throughput
$\Gamma_{B}\left(\mathbf{q}\right)$ of Scheme B at high SNR is given
by
\begin{equation}
\Gamma_{B}\left(\mathbf{q}\right)\rightarrow\frac{W\tilde{R}_{b}\tilde{R}_{d}}{4\tilde{R}_{b}+9Q_{\mathbf{q}}^{B}\tilde{R}_{d}},\:\textrm{as}\: P\rightarrow0.\label{eq:TAqhP}
\end{equation}

\subsection{Analysis of Cache-induced MIMO Cooperation Gain}

From Theorem \ref{thm:PBTB}, we can obtain the following corollary
which quantifies the cache-induced MIMO cooperation gain $\triangle\Gamma\triangleq\Gamma_{A}\left(\mathbf{q}^{*}\right)-\Gamma_{B}\left(\mathbf{q}^{*}\right)$
under the order-optimal cache content replication vector $\mathbf{q}^{*}$
in (\ref{eq:orderq}).
\begin{cor}
[Cache-induced MIMO cooperation gain]\label{cor:Cache-induced-MIMO-cooperation}The
cache-induced MIMO cooperation gain $\triangle\Gamma$ is bounded
as $\triangle\Gamma_{U}\geq\triangle\Gamma\geq\triangle\Gamma_{L}+O\left(PN_{c}^{-\frac{\alpha-2}{2\left(\alpha-1\right)}}\right)$
with 
\begin{eqnarray}
\triangle\Gamma_{a} & = & \Gamma_{A}^{a}\left(\mathbf{q}^{*}\right)-\Gamma_{B}\left(\mathbf{q}^{*}\right),\: a\in\left\{ L,U\right\} ,\label{eq:Taapprx}
\end{eqnarray}
where $\mathbf{q}^{*}$ is given in (\ref{eq:orderq}). Moreover,
as $P,N_{c}\rightarrow\infty$ such that $PN_{c}^{-\frac{\alpha-2}{2\left(\alpha-1\right)}}\rightarrow0$,
we have $\triangle\Gamma\rightarrow\overline{\triangle\Gamma}$, where
\begin{eqnarray}
\overline{\triangle\Gamma} & = & \frac{4W\tilde{R}_{b}\tilde{R}_{d}\left(1-Q_{\mathbf{q}^{*}}^{d}\right)\tilde{R}_{b}}{\left(4Q_{\mathbf{q}^{*}}^{d}\tilde{R}_{b}+9Q_{\mathbf{q}^{*}}^{b}\tilde{R}_{d}\right)\left(4\tilde{R}_{b}+9Q_{\mathbf{q}^{*}}^{B}\tilde{R}_{d}\right)}\label{eq:dltThSNR}\\
 & = & \Theta\left(\frac{\frac{B_{C}}{F}Q_{\mathbf{q}^{*}}^{c}}{\left[\sum_{l=1}^{L}p_{l}^{2/3}\right]^{3}}\right).\nonumber 
\end{eqnarray}

\end{cor}

\begin{figure}
\begin{centering}
\includegraphics[width=80mm]{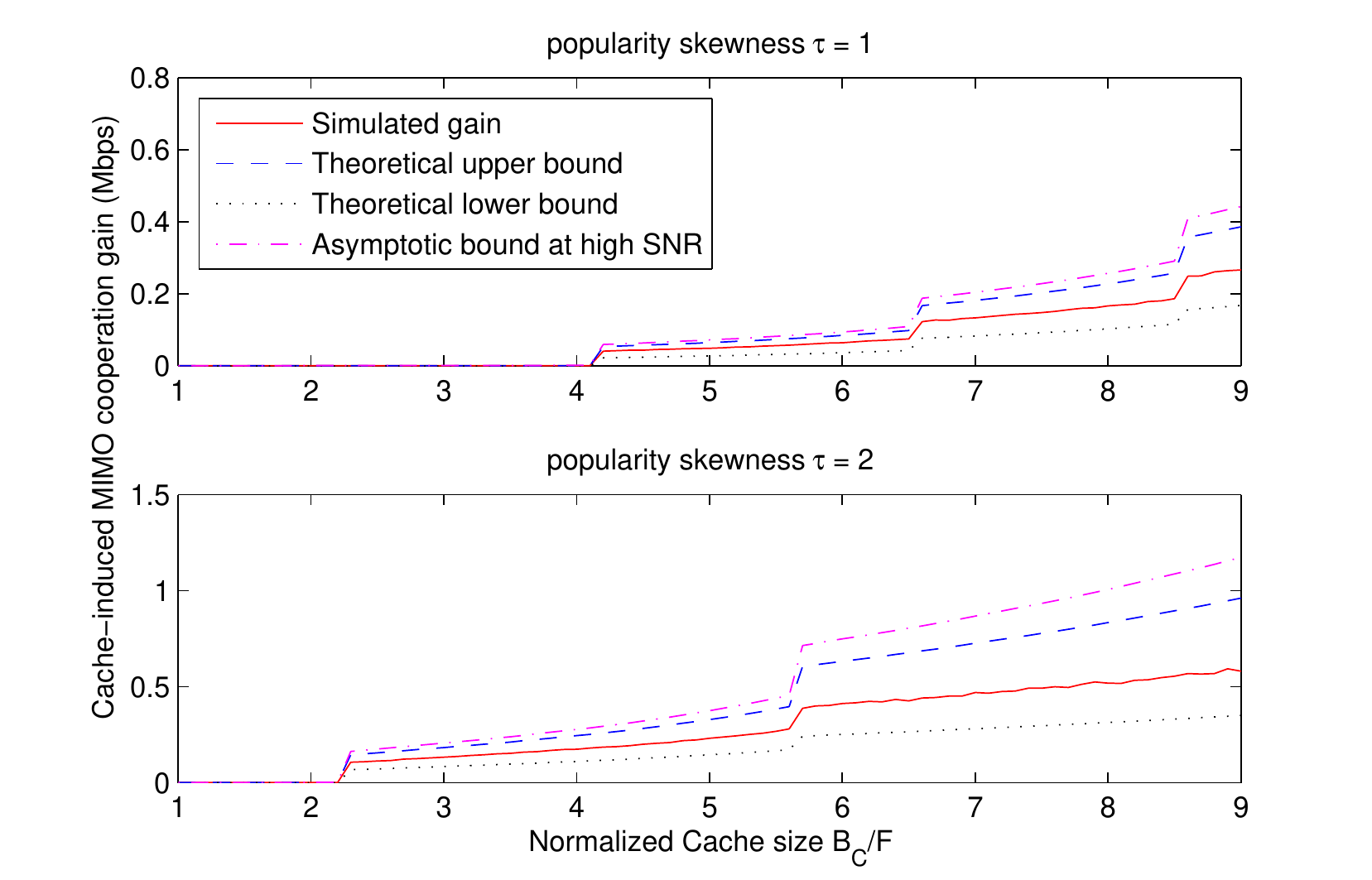}
\par\end{centering}

\protect\caption{\label{fig:VerifyUL}{\small{}Impact of system parameters on the cache-induced
MIMO cooperation gain in a regular C-DWN with $N=144$ BSs, $K=576$
users and $L=10$ files. The system bandwidth is 1MHz and SNR$=20$dB.
The content popularity skewness $\tau=$1 in the upper subplot and
$\tau=$2 in the lower subplot. The Co-MIMO cluster size is $N_{c}=9$.}}
\end{figure}

According to Corollary \ref{cor:Cache-induced-MIMO-cooperation},
the PHY caching gain can be well approximated by $\overline{\triangle\Gamma}$
at high SNR. $\overline{\triangle\Gamma}$ captures the key features
of the actual (simulated) cache-induced MIMO cooperation gain as illustrated
in Fig. \ref{fig:VerifyUL}. From (\ref{eq:dltThSNR}), we have the
following observations. 

\textbf{Impact of the normalized cache size $\tilde{B}_{C}$:} When
$\tilde{B}_{C}<\sum_{l=1}^{L}l^{-\frac{2}{3}\tau}$, we have $q_{l}^{*}<1,\forall l$
and $\overline{\triangle\Gamma}=0$. When $\tilde{B}_{C}\geq\sum_{l=1}^{L}l^{-\frac{2}{3}\tau}$,
$\overline{\triangle\Gamma}>0$ and $\overline{\triangle\Gamma}$
is an increasing function of $\tilde{B}_{C}$, as shown in Fig. \ref{fig:VerifyUL}.
Note that as $\tilde{B}_{C}$ increases, $\overline{\triangle\Gamma}$
has positive jumps because $Q_{\mathbf{q}}^{d}$ is not a continuous
function of $\tilde{B}_{C}$.

\textbf{Impact of the Content Popularity Skewness $\tau$:} As $\tau$
increases, the minimum normalized cache size $\tilde{B}_{C}$ needed
to achieve a non-zero cache-induced MIMO cooperation gain decreases,
and the gain $\overline{\triangle\Gamma}$ also increases for the
same $\tilde{B}_{C}$, as illustrated in Fig. \ref{fig:VerifyUL}.
When $\tau>\frac{3}{2}$, $\sum_{l=1}^{L}p_{l}^{2/3}=\Theta\left(\sum_{l=1}^{L}l^{-\frac{2}{3}\tau}\right)$
is bounded and we can achieve a cache-induced MIMO cooperation gain
of $\Theta\left(1\right)$ even when the normalized cache size $\tilde{B}_{C}=\Theta\left(1\right)$
is fixed and $L\rightarrow\infty$. On the other hand, when $\tau\leq\frac{3}{2}$,
the normalized cache size $\tilde{B}_{C}$ has to increase with $L$
at different orders as $L\rightarrow\infty$ in order to achieve a
significant cache-induced MIMO cooperation gain as shown in the following
corollary.
\begin{cor}
\label{cor:Minimum-cache-size-1}The order of the minimum normalized
cache size $\tilde{B}_{C}^{\textrm{min}}$ needed to achieve a cache-induced
MIMO cooperation gain of $\Theta\left(1\right)$ is the same as that
needed to achieve the linear capacity scaling, as given in Corollary
\ref{cor:Minimum-cache-size}.
\end{cor}

Hence, the minimum required cache size $\tilde{B}_{C}^{\textrm{min}}$
to achieve a large cache-induced MIMO cooperation gain decreases with
the popularity skewness $\tau$, as shown in Fig. \ref{fig:sub-super-critical}.
When either the BS cache size $B_{C}$ is large, or the popularity
skewness $\tau$ is large, the cache-induced MIMO cooperation gain
is significant and it is worthwhile to exploit cache-induced Co-MIMO.

\section{Numerical Results\label{sec:Numerical-Results-and}}

\begin{figure}
\begin{centering}
\textsf{\includegraphics[clip,width=50mm]{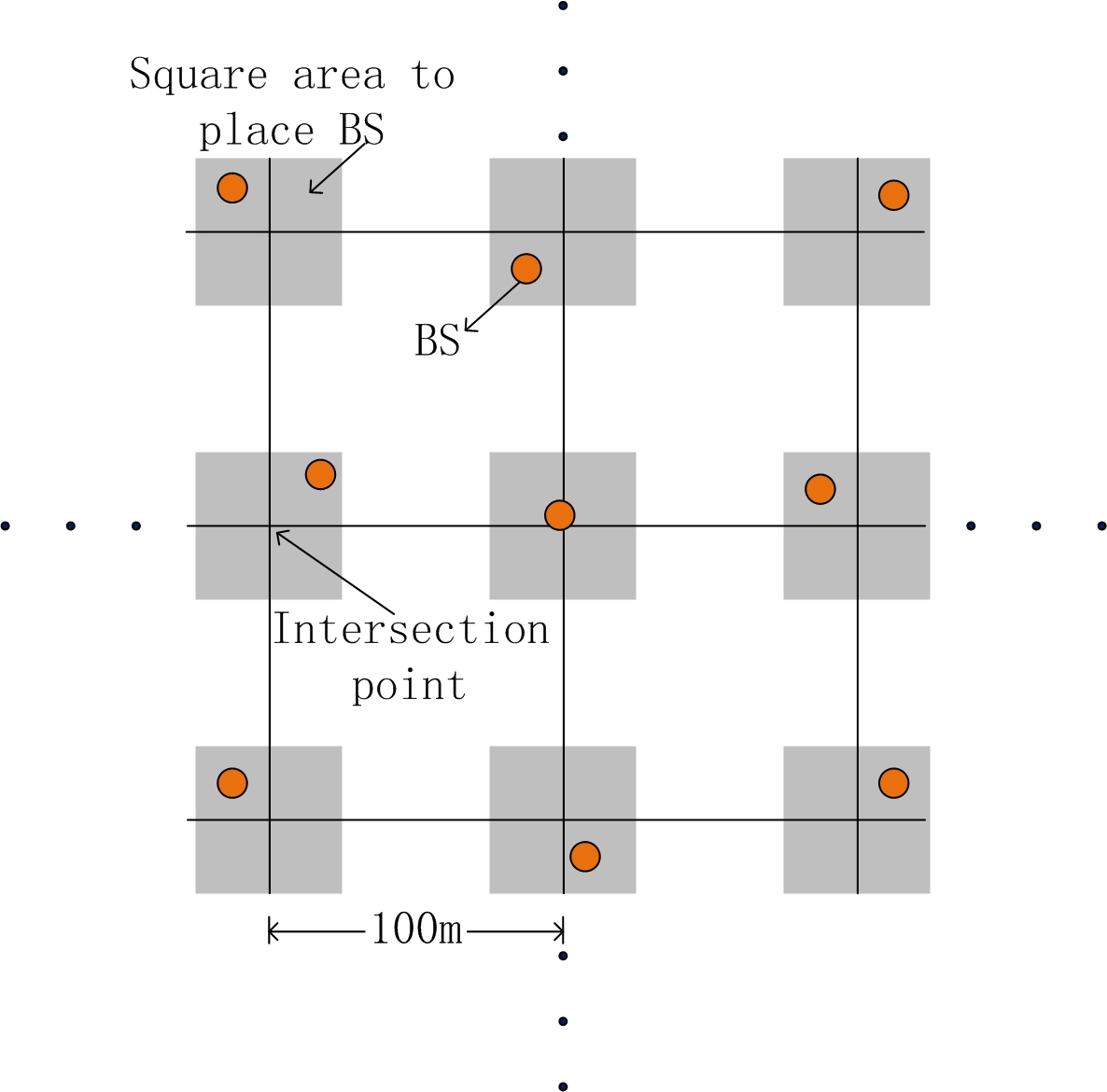}}
\par\end{centering}

\protect\caption{\label{fig:simtop}{\small{}An illustration of BS placement in the
C-DWN network considered in the simulations.}}
\end{figure}

In this section, we illustrate the PHY caching gains for a general
C-DWN network with 225 BSs and 900 users. The locations of the BSs
and users are randomly generated according to Assumption \ref{asm:Place}
with parameters $r_{\textrm{min}}=50$m, $r_{\textrm{max}}=75\sqrt{2}$m,
$d_{\textrm{min}}=10$m and $k_{\textrm{max}}=8$. Specifically, we
first generate a $15\times15$ grid as illustrated in Fig. \ref{fig:simtop}.
Then each BS is randomly placed in the square of side length $50$m
(gray squares in Fig. \ref{fig:simtop}) centered at each of the 225
intersection points on the gird. Finally, the users are placed one
by one in the network. When placing the $k$-th user, we first randomly
pick a cell that has less than $k_{\textrm{max}}=8$ users. Then,
user $k$ is randomly placed at a point within the cell that is at
least $d_{\textrm{min}}=10$m away from the BS in this cell. Only
$N_{0}=10$ BSs have wired backhaul. The system bandwidth is 1MHz.
There are $L=50$ content files on content server and the size of
each file is 1GB. We assume Zipf popularity distribution with different
values $\tau$. The BS cluster size in cache-induced Co-MIMO transmission
is set as $N_{c}=9$.

\begin{figure}
\begin{centering}
\includegraphics[width=80mm]{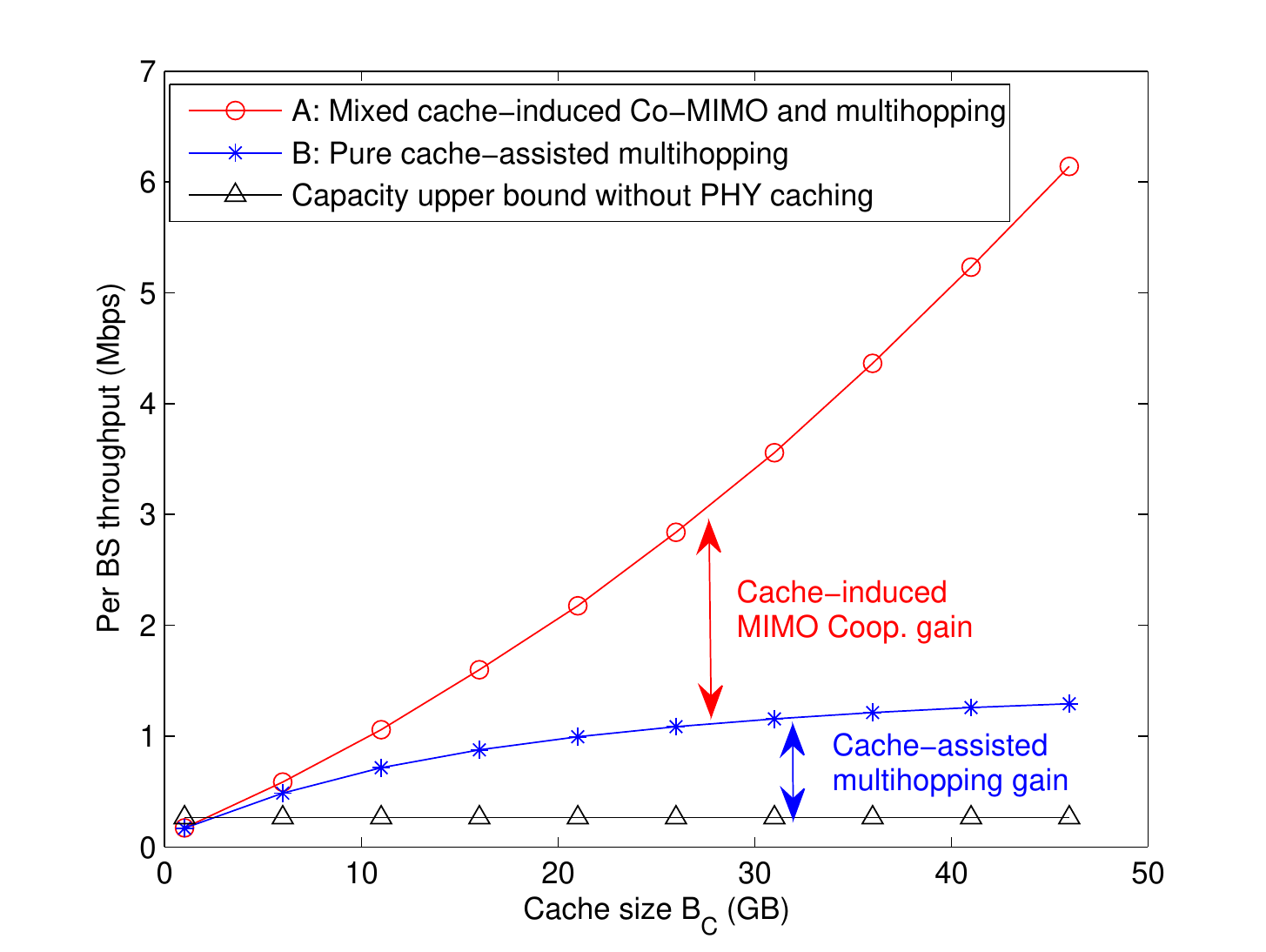}
\par\end{centering}

\protect\caption{\label{fig:GainBC}{\small{}Per BS throughput versus the cache size
$B_{C}$. The content popularity skewness $\tau$ is fixed as 1.5.}}
\end{figure}

In Fig. \ref{fig:GainBC}, we plot the per BS throughput versus the
cache size $B_{C}$. The content popularity skewness $\tau$ is fixed
as 1.5. It can be seen that both the cache-assisted multihopping gain
and cache-induced MIMO cooperation gain increase with the cache size
$B_{C}$, and both gains of PHY caching becomes significant when the
BS cache size is large. 

\begin{figure}
\begin{centering}
\includegraphics[width=80mm]{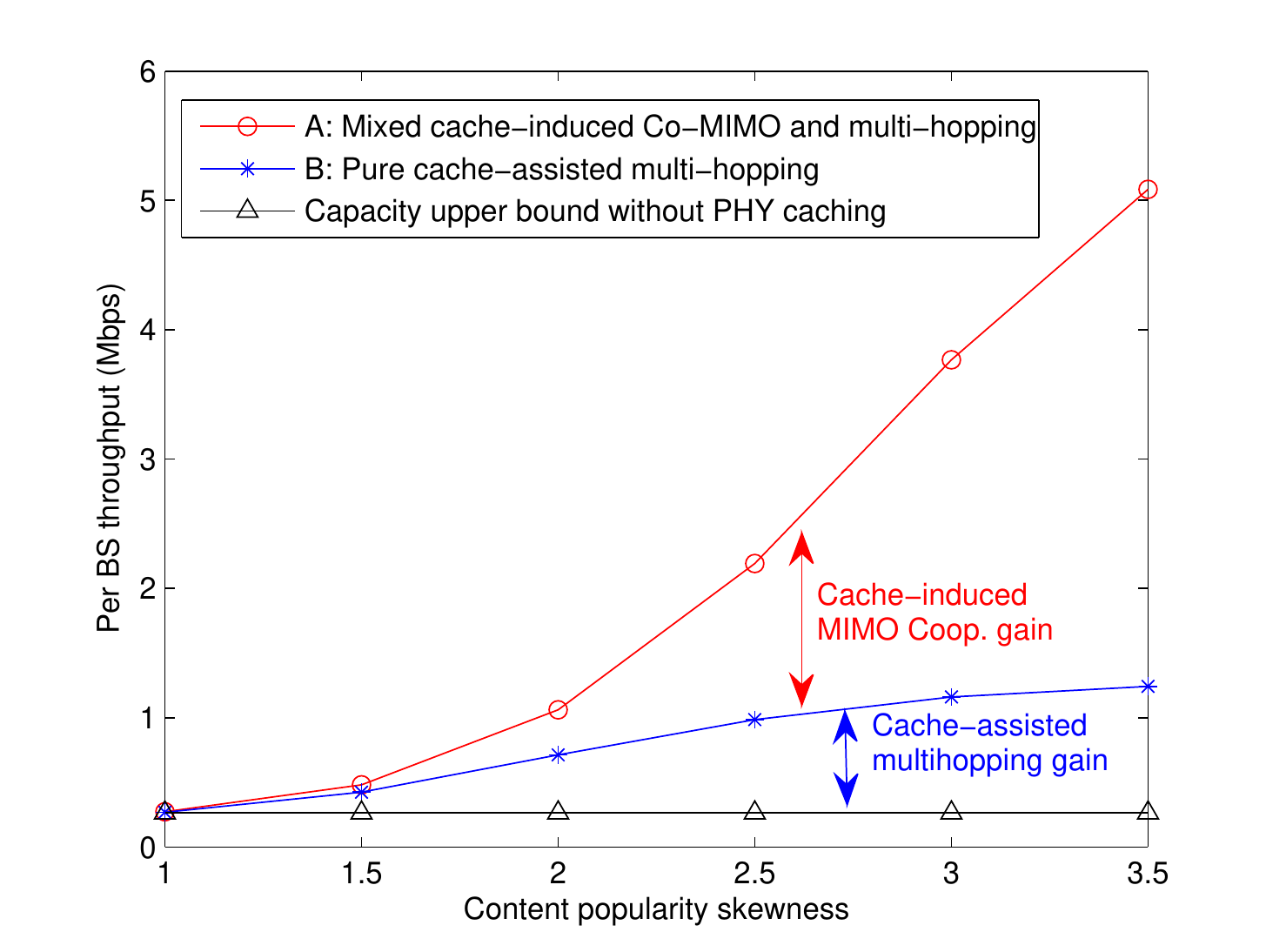}
\par\end{centering}

\protect\caption{\label{fig:Gaintau}{\small{}Per BS throughput versus the popularity
skewness $\tau$. The BS cache size is fixed as 5GB and $L=50$.}}
\end{figure}

We then simulate the case when the BS cache size is much smaller than
the total content size. In Fig. \ref{fig:Gaintau}, we plot the per
BS throughput versus the content popularity skewness $\tau$. The
BS cache size is fixed as 5GB. The results in Fig. \ref{fig:Gaintau}
show that both gains of PHY caching increase with the content popularity
skewness $\tau$. Moreover, even when $B_{C}$ is small compared to
the total content size, it is still possible to achieve a large PHY
caching gain when $\tau$ is large, as shown in Fig. \ref{fig:Gaintau}. 

\section{Discussion\label{sec:Discussion}}

In this section, we extend the results in this paper to study the
impact of network deployment on the throughput scaling laws. Specifically,
we consider the impact of two network parameters, namely the number
of backhaul-connected BSs $N_{0}$ and the system loading in the network.

The number of backhaul-connected BSs $N_{0}$ controls the tradeoff
between performance and deployment cost. In this paper, we focus on
the backhaul-limited case when $N_{0}=\Theta\left(1\right)$. However,
the results can be extended to the case with arbitrary $N_{0}\in\left[0,N\right]$.
In this case, the achievable per user throughput scales according
to $\Theta\left(\max\left\{ R_{\textrm{BL}},\frac{N_{0}}{K}\right\} \right)$,
where $R_{\textrm{BL}}$ is the achievable per user throughput when
$N_{0}=\Theta\left(1\right)$ and its scaling law is described in
Theorem 4, and $\frac{N_{0}}{K}$ is the contribution of the wired
payload backhauls to the throughput scaling law. When $N_{0}=O\left(\frac{K}{\sqrt{LF/B_{C}}}\right)$,
the number of wired payload backhauls is too small to affect the scaling
law. However, as $N_{0}$ increases, the wired payload backhauls may
contribute to the scaling law, depending on the popularity skewness
$\tau$. For example, when the popularity skewness $1<\tau<3/2$ and
$\Theta\left(N_{0}\right)>\Theta\left(\frac{B_{C}K}{FL^{3/2-\tau}}\right)$,
the throughput scales according to $\Theta\left(\frac{N_{0}}{K}\right)$,
which has order-wise improvement compared to $R_{\textrm{BL}}$. On
the other hand, when $\tau>3/2$, the linear capacity scaling law
can be achieved purely by PHY caching without any wired payload backhaul.

The system loading is another important network parameter that may
affect the throughput scaling law. In practice, there are usually
a large number of users within the coverage area of a network, but
not all of them are active. For example, at peak hours, most of the
users will be active, which corresponds to a large system loading,
while at off-peak hours, the ratio of active users will be small,
which corresponds to a light system loading. Hence, the system loading
can be measured using the ratio of active users $\beta\in\left[0,1\right]$.
In this paper, we focus on the more challenging case of full system
loading (i.e., $\beta=1$), where all $K=\Theta\left(N\right)$ users
are assumed to be active. However, the results can be extended to
the case with arbitrary $\beta\in\left[0,1\right]$. In this case,
the achievable per user throughput scales according to $\Theta\left(\min\left\{ \frac{1}{\beta}R_{\textrm{BL}},1\right\} \right)$,
where $R_{\textrm{BL}}$ is the achievable per user throughput under
full system loading $\beta=1$ and its scaling law is described in
Theorem 4. Hence, as $\beta$ decreases, the per user throughput will
increase. However, when $\beta$ is order-wise smaller than a critical
system loading $\beta_{0}=\Theta\left(R_{\textrm{BL}}\right)$, the
order of the aggregate network throughput will degrade compared to
the case when $\beta=1$. The critical system loading $\beta_{0}$
increases with both the cache size $B_{C}$ and popularity skewness
$\tau$.

\section{Conclusion\label{sec:Conclusion}}

In this paper, we propose a PHY caching scheme to address the interference
issue (via cache-induced opportunistic Co-MIMO) and the backhaul cost
issue (via cache-assisted multihopping) in dense wireless networks.
We establish the linear capacity scaling law and present order-optimal
PHY caching and transmission schemes in the backhaul-limited C-DWN.
We further study the impact of various system parameters on the PHY
caching gain and provide fundamental design insight for the backhaul-limited
C-DWN. Specifically, the analytical results show that the minimum
cache size needed to achieve the linear capacity scaling and significant
cache-induced MIMO cooperation gain decreases with the popularity
skewness $\tau$, which measures the concentration of the popularity
distribution. In practice, the popularity skewness $\tau$ can be
large especially for mobile applications \cite{Yamakami_PDCAT06_Zipflaw}.
Hence both benefits of cache-assisted multihopping and cache-induced
Co-MIMO provided by the PHY caching are very effective ways of enhancing
the capacity of dense wireless networks.

\appendix

\subsection{Proof of Lemma \ref{lem:cutB1} and Theorem \ref{thm:CUB-1}\label{sub:Proof-of-CutB}}

The throughput bound in (\ref{eq:Cutb-1}) follows directly from the
cut set bound. Using Assumption \ref{asm:Place}-1), it can be shown
that both $\sum_{n^{'}\notin\mathcal{B}_{P}}G^{b}\left(r_{n^{'},n}^{b}\right)^{-\alpha}$
and $\sum_{k=1}^{K}G^{d}\left(r_{k,n}^{d}\right)^{-\alpha}$ can be
bounded by some constant for all $n\in\mathcal{B}_{P}$. Hence, we
have $\tilde{b}_{U}=\Theta\left(N_{0}\right)$. This completes the
proof of Lemma \ref{lem:cutB1}.

The maximization problem in (\ref{eq:Cutb-1}) is equivalent to:
\begin{eqnarray}
\max_{\left\{ x_{n}\left(\tilde{\mathbf{H}}\right)\right\} }\textrm{E}\left[\sum_{n=1}^{N_{0}}\log\left(1+y_{n}\left(\tilde{\mathbf{H}}\right)x_{n}\left(\tilde{\mathbf{H}}\right)\right)\right]\label{eq:ParChannel}\\
\textrm{s.t.}\:\textrm{E}\left[\sum_{n=1}^{N_{0}}x_{n}\left(\tilde{\mathbf{H}}\right)\right]\leq N_{0}P,\nonumber 
\end{eqnarray}
where $y_{n}\left(\tilde{\mathbf{H}}\right)$ is the $n$-th eigenvalue
of $\tilde{\mathbf{H}}^{\dagger}\tilde{\mathbf{H}}$ and $x_{n}\left(\tilde{\mathbf{H}}\right)$
is the transmit power for the $n$-th eigenchannel. Since $\sum_{n=1}^{N_{0}}y_{n}\left(\tilde{\mathbf{H}}\right)=\textrm{Tr}\left(\tilde{\mathbf{H}}\tilde{\mathbf{H}}^{\dagger}\right)=\tilde{b}_{U},\forall\tilde{\mathbf{H}}$,
the optimal objective value of (\ref{eq:ParChannel}) is upper bounded
by that of
\begin{eqnarray}
\max_{\left\{ x_{n}\left(\tilde{\mathbf{H}}\right),y_{n}\left(\tilde{\mathbf{H}}\right)\right\} }\textrm{E}\left[\sum_{n=1}^{N_{0}}\log\left(1+y_{n}\left(\tilde{\mathbf{H}}\right)x_{n}\left(\tilde{\mathbf{H}}\right)\right)\right]\label{eq:ParChannel-1}\\
\textrm{s.t.}\:\textrm{E}\left[\sum_{n=1}^{N_{0}}x_{n}\left(\tilde{\mathbf{H}}\right)\right]\leq N_{0}P,\:\textrm{E}\left[\sum_{n=1}^{N_{0}}y_{n}\left(\tilde{\mathbf{H}}\right)\right]\leq\tilde{b}_{U},\nonumber 
\end{eqnarray}
where $y_{n}\left(\tilde{\mathbf{H}}\right)$ is also treat as an
optimization variable. When $y_{n}\left(\tilde{\mathbf{H}}\right)$
is fixed as the $n$-th eigenvalue of $\tilde{\mathbf{H}}^{\dagger}\tilde{\mathbf{H}}$,
problem (\ref{eq:ParChannel-1}) reduces to problem (\ref{eq:ParChannel}).
It can be shown that the optimal objective value of (\ref{eq:ParChannel-1})
is equal to that of
\begin{eqnarray}
\max_{\left\{ x_{n}\left(\tilde{\mathbf{H}}\right)\right\} }\textrm{E}\left[\sum_{n=1}^{N_{0}}\log\left(1+\xi x_{n}^{2}\left(\tilde{\mathbf{H}}\right)\right)\right]\label{eq:ParChannel-2}\\
\textrm{s.t.}\:\textrm{E}\left[\sum_{n=1}^{N_{0}}x_{n}\left(\tilde{\mathbf{H}}\right)\right]\leq N_{0}P.\nonumber 
\end{eqnarray}
Since $f\left(\xi,x\right)$ is concave w.r.t. $x$ and $f\left(\xi,x\right)\geq\log\left(1+\xi x^{2}\right),\forall x\geq0$,
the optimal objective value of (\ref{eq:ParChannel-2}) is upper bounded
by that of the following convex problem:
\begin{eqnarray}
\max_{\left\{ x_{n}\left(\tilde{\mathbf{H}}\right)\right\} }\textrm{E}\left[\sum_{n=1}^{N_{0}}f\left(\xi,x_{n}\left(\tilde{\mathbf{H}}\right)\right)\right]\label{eq:Parcvxcha}\\
\textrm{s.t.}\:\textrm{E}\left[\sum_{n=1}^{N_{0}}x_{n}\left(\tilde{\mathbf{H}}\right)\right]\leq N_{0}P.\nonumber 
\end{eqnarray}
By Jensen's inequality, we have 
\begin{eqnarray*}
\textrm{E}\left[\sum_{n=1}^{N_{0}}f\left(\xi,x_{n}\left(\tilde{\mathbf{H}}\right)\right)\right] & \leq & \sum_{n=1}^{N_{0}}f\left(\xi,\textrm{E}\left[x_{n}\left(\tilde{\mathbf{H}}\right)\right]\right)\\
 & \leq & N_{0}f\left(\xi,\frac{1}{N_{0}}\textrm{E}\left[\sum_{n=1}^{N_{0}}x_{n}\left(\tilde{\mathbf{H}}\right)\right]\right).
\end{eqnarray*}
Since $\textrm{E}\left[\sum_{n=1}^{N_{0}}x_{n}\left(\tilde{\mathbf{H}}\right)\right]\leq N_{0}P$,
the optimal objective value of (\ref{eq:Parcvxcha}) is upper bounded
by $N_{0}f\left(\xi,P\right)$. This completes the proof of Theorem
\ref{thm:CUB-1}.

\subsection{Proof of Lemma \ref{lem:SBSbound}-\ref{lem:bounJ}\label{sub:Proof-of-LemmaCM}}

\subsubsection*{Proof of Lemma \ref{lem:SBSbound} }

Let $\mathcal{O}_{k}$ denote a disk centered at the BS $b_{k}$ with
radius $r_{k}^{*}-r_{\textrm{max}}$ and let $\mathcal{O}_{k}^{'}$
denote the intersection of $\mathcal{O}_{k}$ and the network coverage
area (i.e., the square of area $Nr_{0}^{2}$). By Assumption \ref{asm:Place}-2),
any point inside $\mathcal{O}_{k}^{'}$ must lie in the coverage area
of the BSs in $\overline{\mathcal{B}}_{k}$. Since the coverage area
of each BS is less than $\pi r_{\textrm{max}}^{2}$, we must have
$\mathcal{A}\left(\mathcal{O}_{k}^{'}\right)\leq\left|\overline{\mathcal{B}}_{k}\right|\pi r_{\textrm{max}}^{2}$,
where $\mathcal{A}\left(\mathcal{O}_{k}^{'}\right)$ denotes the area
of $\mathcal{O}_{k}^{'}$. Since $\mathcal{A}\left(\mathcal{O}_{k}^{'}\right)\geq\frac{\pi\left(r_{k}^{*}-r_{\textrm{max}}\right)^{2}}{4}$
and $\left|\overline{\mathcal{B}}_{k}\right|\leq\left\lceil 1/q_{l}\right\rceil -1$,
we have $\frac{\pi\left(r_{k}^{*}-r_{\textrm{max}}\right)^{2}}{4}\leq\left(\left\lceil 1/q_{l}\right\rceil -1\right)\pi r_{\textrm{max}}^{2}$
and thus $r_{k}^{*}\leq\left(2\sqrt{\left\lceil 1/q_{l}\right\rceil -1}+1\right)r_{\textrm{max}}$.

\subsubsection*{Proof of Lemma \ref{lem:FRboundM} }

Construct a graph where the vertices are the BS nodes. There is an
edge between any two BSs with distance no more than $r_{I}^{b}$.
Then finding a frequency reuse scheme which satisfies Condition \ref{cond:FR}
is essentially a vertex coloring problem in graph theory. It is well
known that a graph of degree no more than $D_{G}$ can have its vertices
colored by using no more than $D_{G}+1$ colors, with no two neighboring
vertices having the same color \cite{Bondy_Elsevier76_GraphTheory}.
One can therefore allocate the cells with no more than $D_{G}+1$
subbands to satisfies Condition \ref{cond:FR}. The rest is to bound
$D_{G}$, which is the number of BSs in a circle with radius $r_{I}^{b}$.
Using Assumption \ref{asm:Place}-1), we must have $\frac{D_{G}\pi r_{\textrm{min}}^{2}}{4}\leq\pi\left(r_{I}^{b}+\frac{r_{\textrm{min}}}{2}\right)^{2}$.

\subsubsection*{Proof of Lemma \ref{lem:bounJ} }

The routing line segments of a user $k$ can intersect cell $n$ only
when the associated BS $b_{k}$ is within the radius $r_{k}^{*}$.
By lemma \ref{lem:SBSbound}, $r_{k}^{*}\leq\left(2\sqrt{\left\lceil 1/q_{l}\right\rceil -1}+1\right)r_{\textrm{max}}$
for any user $k$ requesting the $l$-th file. Let $n_{l}^{\textrm{max}}$
denote the maximum number of BSs in a circle with radius $\left(2\sqrt{\left\lceil 1/q_{l}\right\rceil -1}+1\right)r_{\textrm{max}}$.
Then according to the above analysis, the average number of users
whose routing line segments intersect cell $n$ is upper bounded by
$\sum_{l}p_{l}n_{l}^{\textrm{max}}k_{\textrm{max}}$. Using similar
analysis as for $D_{G}$, it can be shown that $n_{l}^{\textrm{max}}\leq\left(\frac{\left(4\sqrt{\left\lceil 1/q_{l}\right\rceil -1}+2\right)r_{\textrm{max}}}{r_{\textrm{min}}}+1\right)^{2}$.
This completes the proof.

\subsection{Proof of Theorem \ref{thm:CLB}\label{sub:Proof-of-TheoremCLB}}

Clearly, the throughput of each wireless inter-BS link $C_{b}=\Theta\left(1\right)$,
and the per cell downlink access transmission rate $C_{d}=\Theta\left(1\right)$.
Suppose we want to support a per user throughput of $R$. Then the
traffic to be relayed by a BS due to wireless inter-BS transmission
is upper bounded by $JR$ according to Lemma \ref{lem:bounJ} and
the traffic to be handled by a BS due to downlink access transmission
is upper bounded by $k_{\textrm{max}}R$. Clearly, for a user requesting
a file with multihop cache mode, a throughput of $R$ can be supported
by Scheme A if no BS is overloaded, i.e., $JR\leq C_{b}$ and $k_{\textrm{max}}R\leq C_{d}$.
Hence, a per user throughput of $R=\min\left(\frac{C_{b}}{J},\frac{C_{d}}{k_{\textrm{max}}}\right)$
is achievable. Since $J=\Theta\left(\sum_{l=1}^{L}p_{l}\sqrt{\frac{1}{q_{l}}}\right)$
from Lemma \ref{lem:bounJ}, the order of $R$ is given by (\ref{eq:Rorder}).
On the other hand, for a user requesting a file with Co-MIMO cache
mode, it is clear that a throughput of $\Theta\left(1\right)\geq\Theta\left(\frac{1}{J}\right)$
is achievable. As a result, an overall per user throughput of $R=\Theta\left(\frac{1}{J}\right)$
is also achievable.

Consider the convex problem of maximizing the order of per user throughput:
\begin{equation}
\min_{\mathbf{q}}\sum_{l=1}^{L}p_{l}\sqrt{\frac{1}{q_{l}}},\: s.t.\: q_{l}\in\left[0,1\right],\forall l,\:\sum_{l=1}^{L}q_{l}\leq\frac{B_{C}}{F}.\label{eq:optq}
\end{equation}
By analyzing the KKT conditions, it can be shown that (\ref{eq:orderq})
is an order optimal solution for (\ref{eq:optq}). By substituting
(\ref{eq:optq}) into $R=\Theta\left(1/\sum_{l=1}^{L}p_{l}\sqrt{\frac{1}{q_{l}}}\right)$,
it can be verified that the order of $R$ is the same as that in Theorem
\ref{thm:scalingLinf}.

\subsection{Proof of Lemma \ref{lem:RbA}\label{sub:Proof-of-LemmaRbARdA} }

For each wireless inter-BS link, the bandwidth is $\frac{W_{b}}{36}$
(note that each BS has four wireless inter-BS links) and the transmit
power on this bandwidth is $\frac{W_{b}P}{36W^{'}}$. The noise power
is $\frac{W_{b}\eta_{0}}{36}$ and it can be verified that the interference
power is $\frac{W_{b}P}{36W^{'}}G^{b}I_{R}\left(1,3\right)$. Hence
the SINR of each wireless inter-BS link is $\frac{PG^{b}r_{0}^{-\alpha}}{W^{'}\eta_{0}+PG^{b}I_{R}\left(1,3\right)}$
and the rate is given by $W_{b}R_{b}^{B}$. 

Similarly, for downlink access transmission, the bandwidth allocated
to each cell is $\frac{W_{d}}{4}$ and the transmit power on this
bandwidth is $\frac{W_{d}P}{4W^{'}}$. The noise power is $\frac{W_{d}\eta_{0}}{4}$
and it can be verified that the interference power is $\frac{W_{d}P}{4W^{'}}G^{d}I_{R}\left(\frac{d_{0}}{r_{0}},2\right)$.
Hence the SINR in downlink access transmission is $\frac{PG^{d}d_{0}^{-\alpha}}{W^{'}\eta_{0}+PG^{d}I_{R}\left(\frac{d_{0}}{r_{0}},2\right)}$
and the per user average downlink access transmission rate is given
by $W_{d}R_{d}^{B}$.

\subsection{Proof of Theorem \ref{lem:LoCoMIMObound}\label{sub:Proof-of-TheoremCoMIMO}}

Consider the dual uplink system \cite{Liu_10sTSP_Fairness_rate_polit_WF}
of the downlink system where in each cluster, the scheduled users
act as the transmitters, the BSs act as the receivers, and the uplink
channels are the Hermition of the corresponding downlink channels.
We first study the per cluster throughput of the uplink system. Then
the results can be transferred to the downlink system using the downlink-uplink
duality \cite{Liu_10sTSP_Fairness_rate_polit_WF}. At each time slot,
the BS clusters are randomly formed and each cluster contains $N_{c}$
BSs in a square. Consider the following achievable scheme for the
uplink system. In each cluster, the users whose distance from the
cluster boundary is less than a threshold $d_{b}=\Theta\left(\left(N_{c}r_{0}\right)^{\frac{1}{2\left(\alpha-1\right)}}\right)$
is not allowed to transmit for interference control. The other $\overline{N}_{c}=\Theta\left(N_{c}-N_{c}^{\frac{\alpha}{2\left(\alpha-1\right)}}\right)$
scheduled users transmit at a constant power $P^{'}=\frac{36W_{c}P}{4W_{b}+9W_{d}+36W_{c}}$.
Consider a reference cluster. Treating the interference from other
clusters as noise, we can achieve a per cluster throughput of
\[
C_{u}=W_{c}\textrm{E}\left[\log\left|\mathbf{I}+P^{'}\mathbf{\Omega}^{-1}\mathbf{H}_{c}\mathbf{H}_{c}^{\dagger}\right|\right],
\]
where $\mathbf{H}_{c}=\left[h_{i,j}\right]_{i=1,...,N_{c},j=1,...,\overline{N}_{c}}\in\mathbb{C}^{N_{c}\times\overline{N}_{c}}$
and $h_{i,j}$ is the uplink channel between the $j$-th scheduled
user and the $i$-th BS in the reference cluster, $\mathbf{\Omega}$
is the covariance of the inter-cluster interference plus noise at
the BSs. Using Jensen's inequality, we have
\begin{eqnarray*}
C_{u} & \geq & W_{c}\textrm{E}\left[\log\left|\mathbf{I}+P^{'}\textrm{E}\left[\mathbf{\Omega}|\mathbf{H}_{c}\right]^{-1}\mathbf{H}_{c}\mathbf{H}_{c}^{\dagger}\right|\right]\\
 & = & W_{c}\textrm{E}\left[\log\left|\mathbf{I}+\frac{P^{'}}{W_{c}\eta_{0}+\Theta\left(Pd_{b}^{2-\alpha}\right)}\mathbf{H}_{c}\mathbf{H}_{c}^{\dagger}\right|\right]\\
 & = & W_{c}\textrm{E}\left[\log\left|\mathbf{I}+\frac{P^{'}}{W_{c}\eta_{0}}\mathbf{H}_{c}^{\dagger}\mathbf{H}_{c}\right|\right]+\Theta\left(PN_{c}^{\frac{\alpha}{2\left(\alpha-1\right)}}\right),
\end{eqnarray*}
where the first equality follows from the fact that $\textrm{E}\left[\mathbf{\Omega}|\mathbf{H}_{c}\right]$
is diagonal%
\footnote{This is because the channel coefficients of the cross links from different
users in other clusters have independent distributions with zero means.%
} and the nearest interfering user is at least $d_{b}$ away from the
reference BSs. 

We can use the same technique as in Appendix I of \cite{Tse_IT07_CapscalingHMIMO}
to bound the term $\overline{C}_{u}\triangleq W_{c}\textrm{E}\left[\log\left|\mathbf{I}+\frac{P^{'}}{W_{c}\eta_{0}}\mathbf{H}_{c}^{\dagger}\mathbf{H}_{c}\right|\right]$.
Let $\lambda$ be chosen uniformly among the $\overline{N}_{c}$ eigenvalues
of $\frac{\mathbf{H}_{c}^{\dagger}\mathbf{H}_{c}}{\overline{N}_{c}}$.
Then
\begin{eqnarray*}
\overline{C}_{u} & \geq & W_{c}\overline{N}_{c}\textrm{E}\left[\log\left(1+\frac{\overline{N}_{c}P^{'}}{W_{c}\eta_{0}}\lambda\right)\right]\\
 & \geq & W_{c}\overline{N}_{c}\log\left(1+\frac{\overline{N}_{c}P^{'}}{W_{c}\eta_{0}}t\right)\Pr\left(\lambda>t\right)
\end{eqnarray*}
for any $t\geq0$. By the Paley-Zygmund inequality, we have 
\[
\Pr\left(\lambda>t\right)\geq\frac{\left(\textrm{E}\left(\lambda\right)-t\right)^{2}}{\textrm{E}\left(\lambda^{2}\right)},\:0\leq t<\textrm{E}\left(\lambda\right).
\]
Following similar analysis as in Appendix I of \cite{Tse_IT07_CapscalingHMIMO},
we have 
\[
\textrm{E}\left(\lambda\right)=\frac{G^{d}}{\overline{N}_{c}^{2}}\sum_{k=1}^{\overline{N}_{c}}\sum_{i=1}^{N_{c}}r_{i,k}^{-\alpha}=\frac{G^{d}\left(G_{C}+O\left(N_{c}^{-\frac{\alpha-2}{2\left(\alpha-1\right)}}\right)\right)}{\overline{N}_{c}},
\]
where the last equality follows from $\sum_{i=1}^{N_{c}}r_{i,k}^{-\alpha}=G_{C}+O\left(N_{c}^{-\frac{\alpha-2}{2\left(\alpha-1\right)}}\right)$,
and
\begin{eqnarray*}
\textrm{E}\left(\lambda^{2}\right) & = & \frac{2\left(G^{d}\right)^{2}}{\overline{N}_{c}^{3}}\sum_{k=1}^{\overline{N}_{c}}\sum_{i=1}^{N_{c}}r_{i,k}^{-\alpha}\sum_{l=1}^{N_{c}}r_{l,k}^{-\alpha}\\
 & = & \frac{2\left(G^{d}\right)^{2}\left(G_{C}+O\left(N_{c}^{-\frac{\alpha-2}{2\left(\alpha-1\right)}}\right)\right)^{2}}{\overline{N}_{c}^{2}}.
\end{eqnarray*}
Choose $t=\frac{G^{d}d_{0}^{-\alpha}}{\overline{N}_{c}}$. We have
\[
\overline{C}_{u}\geq W_{c}\overline{N}_{c}\rho\log\left(1+\frac{P^{'}G^{d}d_{0}^{-\alpha}}{W_{c}\eta_{0}}\right)+O\left(PN_{c}^{\frac{\alpha}{2\left(\alpha-1\right)}}\right).
\]

According to the downlink-uplink duality \cite{Liu_10sTSP_Fairness_rate_polit_WF},
a per cluster throughput of $C_{d}=C_{u}\geq\overline{C}_{u}+\Theta\left(PN_{c}^{\frac{\alpha}{2\left(\alpha-1\right)}}\right)$
can be achieved with equal or less total network power. Since the
BS clusters are randomly formed, all users and BSs are statistically
symmetric. As a result, the average downlink access transmission rate
$R_{d}^{C'}$ of a user on the Co-MIMO band is lower bounded as 
\begin{eqnarray*}
R_{d}^{C'} & \geq & \frac{C_{d}}{4N_{c}}\\
 & \geq & \frac{\rho W_{c}}{4}\log\left(1+\frac{P^{'}G^{d}d_{0}^{-\alpha}}{W_{c}\eta_{0}}\right)+\Theta\left(PN_{c}^{-\frac{\alpha-2}{2\left(\alpha-1\right)}}\right)\\
 & \geq & \frac{\rho W_{c}}{4}\log\left(1+\frac{PG^{d}d_{0}^{-\alpha}}{W\eta_{0}}\right)+\Theta\left(PN_{c}^{-\frac{\alpha-2}{2\left(\alpha-1\right)}}\right)
\end{eqnarray*}
and the average power at each BS required to achieve the above per
user rate is no more than $P^{'}$, where the last inequality follows
from $\frac{P^{'}}{W_{c}}=\frac{36P}{4W_{b}+9W_{d}+36W_{c}}\geq\frac{P}{W}$.
On the other hand, using the cut set bound between all BSs and a user,
we have
\begin{equation}
R_{d}^{C'}\leq\frac{W_{c}}{4}\log\left(1+\frac{P^{'}G^{d}G_{C}}{W_{c}\eta_{0}}\right)\overset{\textrm{a}}{\leq}W_{c}\log\left(1+\frac{9PG^{d}G_{C}}{W\eta_{0}}\right),\label{eq:RcpU}
\end{equation}
where (\ref{eq:RcpU}-a) follows from $\frac{P^{'}}{W_{c}}\leq\frac{9P}{W}$.

\subsection{Proof of Theorem \ref{thm:PBTB}\label{sub:Proof-of-TheoremPBTB}}

First, we derive the maximum supportable per user throughput on the
cache-assisted multihop band (i.e., inter-BS band plus downlink access
band). Suppose the users request files $l\in\Omega\left(\mathbf{q}\right)$
at per user throughput $R$ on the cache-assisted multihop band. Let
us focus on a reference user $k$. According to the proposed source
BS set selection scheme, for each requested file segment, $q_{l}L_{S}$
parity bits are obtained from the associated BS, a total number of
$4mq_{l}L_{S}$ parity bits are obtained from the source BSs in $\mathcal{B}_{k,m}$
for $1\leq m<\phi\left(q_{l}\right)$, and a total number of $\left(1-\left(1+2\phi^{2}\left(q_{l}\right)-2\phi\left(q_{l}\right)\right)q_{l}\right)L_{S}$
parity bits are obtained from the source BSs in $\mathcal{B}_{k,\phi\left(q_{l}\right)}$.
As a result, the wireless inter-BS traffic $T_{l}$ induced by a single
user requesting the $l$-th file is
\begin{eqnarray*}
T_{l} & = & \sum_{m=1}^{\phi\left(q_{l}\right)-1}4m^{2}q_{l}R+\\
 &  & \phi\left(q_{l}\right)\left(1-\left(1+2\phi^{2}\left(q_{l}\right)-2\phi\left(q_{l}\right)\right)q_{l}\right)R\\
 & = & \left(\phi\left(q_{l}\right)\left(1-q_{l}\right)-\frac{2}{3}(\phi^{3}\left(q_{l}\right)-\phi\left(q_{l}\right))q_{l}\right)R.
\end{eqnarray*}
Note that the ratio between the number of wireless inter-BS links
and the number of users is $\lim_{N\rightarrow\infty}\frac{4N-4\sqrt{N}}{4N}=1$.
Since all wireless inter-BS links are symmetric, the total wireless
inter-BS traffic induced by all users are equally partitioned among
all wireless inter-BS links. Hence, the corresponding traffic on each
wireless inter-BS link is $\sum_{l\in\Omega\left(\mathbf{q}\right)}p_{l}T_{l}/Q_{\mathbf{q}}^{d}$
and we must have $\sum_{l\in\Omega\left(\mathbf{q}\right)}p_{l}T_{l}/Q_{\mathbf{q}}^{d}\leq W_{b}R_{b}\left(W_{b},W_{c}\right)$.
Meanwhile. we have $R\leq\left(W-W_{b}-W_{c}\right)R_{d}\left(W_{b},W_{c}\right)$.
Hence, the maximum supportable per user throughput on the cache-assisted
multihop band $R_{m}\left(W_{b},W_{c}\right)=\min\left(\frac{Q_{\mathbf{q}}^{d}W_{b}R_{b}\left(W_{b},W_{c}\right)}{Q_{\mathbf{q}}^{b}},\left(W-W_{b}-W_{c}\right)R_{d}\left(W_{b},W_{c}\right)\right)$.
On the other hand, it is easy to see that the maximum supportable
per user throughput on the Co-MIMO band is $W_{c}R_{c}\left(W_{b},W_{c}\right)$.

The overall per user throughput is defined as $T_{\mathbf{q}}\left(W_{b},W_{c}\right)=\lim_{L_{0}\rightarrow\infty}\frac{L_{0}F}{t_{0}}$,
where $L_{0}$ is the total number files delivered to a reference
user $k$ within time $t_{0}$. Let $L_{l}$ denote the number of
delivering the $l$-th file. Clearly, we have
\[
\lim_{L_{0}\rightarrow\infty}t_{0}=\max\left(\frac{\sum_{l\in\Omega\left(\mathbf{q}\right)}FL_{l}}{R_{m}\left(W_{b},W_{c}\right)},\frac{\sum_{l\in\overline{\Omega}\left(\mathbf{q}\right)}FL_{l}}{W_{c}R_{c}\left(W_{b},W_{c}\right)}\right)
\]
\begin{eqnarray*}
T_{\mathbf{q}}\left(W_{b},W_{c}\right) & = & \lim_{L_{0}\rightarrow\infty}\frac{L_{0}F}{\max\left(\frac{\sum_{l\in\Omega\left(\mathbf{q}\right)}FL_{l}}{R_{m}\left(W_{b},W_{c}\right)},\frac{\sum_{l\in\overline{\Omega}\left(\mathbf{q}\right)}FL_{l}}{W_{c}R_{c}\left(W_{b},W_{c}\right)}\right)}\\
 & = & \lim_{L_{0}\rightarrow\infty}\frac{1}{\max\left(\frac{Q_{\mathbf{q}}^{d}}{R_{m}\left(W_{b},W_{c}\right)},\frac{Q_{\mathbf{q}}^{c}}{W_{c}R_{c}\left(W_{b},W_{c}\right)}\right)},
\end{eqnarray*}
where the last equality follows from $\lim_{L_{0}\rightarrow\infty}\frac{L_{l}}{L_{0}}=p_{l}$.
Clearly, $T_{\mathbf{q}}^{L}\left(W_{b},W_{c}\right)+O\left(PN_{c}^{-\frac{\alpha-2}{2\left(\alpha-1\right)}}\right)\leq T_{\mathbf{q}}\left(W_{b},W_{c}\right)\leq T_{\mathbf{q}}^{U}\left(W_{b},W_{c}\right)$,
where $T_{\mathbf{q}}^{a}\left(W_{b},W_{c}\right)$ is obtained by
replacing $R_{b}\left(W_{b},W_{c}\right),R_{d}\left(W_{b},W_{c}\right),R_{c}\left(W_{b},W_{c}\right)$
in $T_{\mathbf{q}}\left(W_{b},W_{c}\right)$ with $R_{b}^{a},R_{d}^{a},R_{c}^{a}$
for $a\in\left\{ U,L\right\} $. Hence, the per BS throughput $\Gamma_{A}\left(\mathbf{q}\right)$
is bounded as $\Gamma_{A}^{U}\left(\mathbf{q}\right)\geq\Gamma_{A}\left(\mathbf{q}\right)\geq\Gamma_{A}^{L}\left(\mathbf{q}\right)+O\left(PN_{c}^{-\frac{\alpha-2}{2\left(\alpha-1\right)}}\right)$
with $\Gamma_{A}^{a}\left(\mathbf{q}\right)\triangleq\max_{W_{b},W_{c}\in\left(0,W\right)}4T_{\mathbf{q}}^{a}\left(W_{b},W_{c}\right)$
for $a\in\left\{ U,L\right\} $. It can be verified that $\Gamma_{A}^{a}\left(\mathbf{q}\right)$
is given in (\ref{eq:Taqgen}). Finally, as $P,N_{c}\rightarrow\infty$
such that $PN_{c}^{-\frac{\alpha-2}{2\left(\alpha-1\right)}}\rightarrow0$,
we have $R_{b}^{a}\rightarrow\frac{1}{36}\tilde{R}_{b}$, $R_{d}^{a}\rightarrow\frac{1}{16}\tilde{R}_{d},$
$R_{b}^{a}/R_{c}^{a}\rightarrow0$, and $R_{d}^{a}/R_{c}^{a}\rightarrow0$
for $a\in\left\{ L,U\right\} $, from which (\ref{eq:TAq}) follows.


\end{document}